\documentclass[a4paper,11pt,oneside]{article}

\usepackage[english]{babel}
\usepackage[T1]{fontenc}
\usepackage[utf8]{inputenc}

\usepackage[pdftex,unicode]{hyperref}
\hypersetup{pdftitle=An Intraday GARCH Model for Discrete Price Changes and Irregularly Spaced Observations}
\hypersetup{pdfauthor=Vladimír Holý}

\usepackage{xcolor}
\definecolor{mycolor}{rgb}{0.00,0.50,0.67}
\hypersetup{colorlinks=true, linkcolor=mycolor, anchorcolor=mycolor, citecolor=mycolor, filecolor=mycolor, urlcolor=mycolor}

\usepackage[margin=60pt]{geometry}
\usepackage{amsmath}
\usepackage{amssymb}
\usepackage{amsthm}

\usepackage[group-minimum-digits=3]{siunitx}

\usepackage{graphicx}
\usepackage{float}
\usepackage{hhline}
\usepackage{multirow}
\usepackage{booktabs}
\usepackage{dcolumn}

\usepackage[authoryear]{natbib}

\usepackage{pifont}
\newcommand{\cmark}{\color{mycolor}\ding{51}}
\newcommand{\xmark}{\color{gray}\ding{55}}

\begin{document}

\begin{center}
{\Large \bfseries An Intraday GARCH Model for Discrete Price Changes \\ and Irregularly Spaced Observations}
\end{center}

\begin{center}
{\bfseries Vladimír Holý} \\
Prague University of Economics and Business \\
Winston Churchill Square 1938/4, 130 67 Prague 3, Czechia \\
\href{mailto:vladimir.holy@vse.cz}{vladimir.holy@vse.cz} \\
\end{center}

\noindent
\textbf{Abstract:}
We develop a novel observation-driven model for high-frequency prices. We account for irregularly spaced observations, simultaneous transactions, discreteness of prices, and market microstructure noise. The relation between trade durations and price volatility, as well as intraday patterns of trade durations and price volatility, is captured using smoothing splines. The dynamic model is based on the zero-inflated Skellam distribution with time-varying volatility in a score-driven framework. Market microstructure noise is filtered by including a moving average component. The model is estimated by the maximum likelihood method. In an empirical study of the IBM stock, we demonstrate that the model provides a good fit to the data. Besides modeling intraday volatility, it can also be used to measure daily realized volatility.
\\

\noindent
\textbf{Keywords:} Ultra-High-Frequency Data, Trade Duration, Price Volatility, UHF-GARCH Model, Score-Driven Model, Skellam Distribution.
\\

\noindent
\textbf{JEL Classification:} C22, C41, C58, G12.
\\

\section{Introduction}
\label{sec:intro}

Modeling intraday volatility presents several challenges in contrast to modeling volatility at the daily level as high-frequency data have distinct characteristics. A widely used tool for modeling daily volatility is the class of generalized autoregressive conditional heteroskedasticity (GARCH) models with seminal contributions by \cite{Engle1982a}, \cite{Bollerslev1986,Bollerslev1987}, and \cite{Nelson1991}. A variety of intraday GARCH models extending daily models therefore emerged, following the call for research in this direction by \cite{Engle2002b}. In this paper, we focus on the following four characteristics of high-frequency prices in the context of intraday GARCH models:

\textbf{Irregularly spaced observations.} \cite{Engle2000} coined the term ultra-high-frequency (UHF) data, which refer to records of every transaction made resulting in irregularly spaced observations. Such data require special treatment in econometric modeling. \cite{Engle1998} proposed to model times between successive transactions, also known as trade durations, by the autoregressive conditional duration (ACD) model. Furthermore, \cite{Engle2000} proposed to model the variance per time unit using irregularly spaced observations by the UHF-GARCH model. \cite{Ghysels1998} proposed an alternative GARCH model for UHF data in which the total variance is modeled but the GARCH parameters are functions of the expected duration. \cite{Meddahi2006} highlighted the differences between these two models. The UHF-GARCH model of \cite{Engle2000} was further applied e.g.\ by \cite{Racicot2008} and \cite{Huptas2016}.

\textbf{Simultaneous transactions.} A particular issue of UHF data is the occurence of transactions with the same timestamp resulting in zero durations. \cite{Engle1998} considered these transactions to be split transactions which belong to a single trade and decided to aggregate them. Note that zero duration does not necessarily mean zero return as transactions can be executed at the same time at different prices. \cite{Blasques2022a} further studied the issue of zero durations and pointed out that close-to-zero or zero durations (depending on the precision of timestamps) may account for the majority of observations in UHF datasets from recent years. Additionally, a major issue is that zero durations do not equate to split transactions in recent datasets---zero durations could correspond to separate transactions occurring simultaneously, and conversely, split transactions can be executed with delays resulting in nonzero durations. Aggregation is thus not a suitable solution as it does not solve the problem of simultaneous transactions and removes a large portion of observations. However, zero durations need to be addressed because when measuring price variance per time unit, as \cite{Engle2000} did, returns are divided by the square root of the corresponding trade duration. Zero durations with nonzero returns of course disrupt this concept of variance per time unit. 

\textbf{Discretness of prices.} Financial assets are traded on a discrete grid of prices. On the NYSE and NASDAQ exchanges, e.g., stocks are traded with precision to one cent. This discreteness has a large impact on the distribution of returns (see, e.g., \citealp{Munnix2010} for empirical evidence). Consequently, a strand of literature emerged that focuses on dynamic volatility models for discrete price changes based on the Skellam distribution and its generalizations. \cite{Koopman2017a} modified the Skellam distribution by transferring probability mass between 0, 1, and -1 values and used it in a dynamic state space model for price changes. \cite{Koopman2018} took a multidimensional approach and modeled price changes by a score-driven model based on a discrete copula with Skellam margins. \cite{Alomani2018} used the Skellam GARCH model for drug crimes. \cite{Goncalves2020} studied more general integer GARCH processes with applications to polio cases and Olympic medals won. \cite{Cui2021} used a GARCH model based on the Skellam distribution with modified probabilities for daily price changes. \cite{Doukhan2021} studied theoretical properties of integer GARCH processes. \cite{Catania2022} used the zero-inflated Skellam distribution in a hidden Markov model for multivariate price changes. Note that none of these studies utilize UHF data and are limited only to a fixed frequency---e.g., 1 second in \cite{Koopman2017a}, 10 second in \cite{Koopman2018}, and 15 second in \cite{Catania2022}. In contrast to time series models, Skellam models in continuous time were analyzed by \cite{Barndorff-Nielsen2012} and \cite{Shephard2017}. An alternative approach was adopted by \cite{Holy2022b} who modeled prices directly, instead of price changes or logarithmic returns, by the double Poisson distribution.

\textbf{Market microstructure noise.} A well documented feature of high frequency data is market microstructure noise---a deviation from the fundamental efficient price (see, e.g., \citealp{Hansen2006} for an in-depth study). It is caused by price discreteness but also by bid-ask bounce, asymmetric information of traders, and other informational effects. It plays a key role in nonparametric estimation of quadratic variation and integrated variance as it significantly biases realized variance at higher frequencies (see, e.g., \citealp{Holy2023a} for an overview of noise-robust estimators). Regarding parametric processes, independent market microstructure noise induces a moving average component of order one. Specifically, \cite{Ait-Sahalia2005} showed that Wiener process contaminated by independent market microstructure noise sampled at discrete times corresponds to ARIMA(0,1,1) process and \cite{Holy2018d} showed that discretized noisy Ornstein--Uhlenbeck process corresponds to ARIMA(1,0,1) process. 

Table \ref{tab:literature} lists notable high-frequency models and summarizes their features. Note that none of these models address all four of the above high-frequency characteristics.

he goal of this paper is to propose a novel high-frequency model that addresses all four presented characteristics of UHF data, i.e., to combine the UHF-GARCH approach with the Skellam-GARCH approach while accounting for simultaneous transactions and market microstructure noise. Furthermore, we aim to assess the importance of the individual components of the proposed model capturing these four characteristics in comparison to alternative model specifications.

Our approach starts with nonparametric estimation of diurnal trends in trade durations and squared price changes using smoothing splines. When both these time series are adjusted for diurnal trends, their relation is estimated using smoothing splines. Next, we build our dynamic model. The original (unadjusted) price changes are assumed to follow the zero-inflated Skellam distribution of \cite{Skellam1946} with time-varying mean and variance and static probability of zero-inflation. The dynamic mean follows MA(1) process to capture the effects of market microstructure noise. As high-frequency data exhibit zero expected returns, we set the intercept to zero. In the Skellam distribution, the variance is required to be higher than the absolute value of the mean, which is suitable for high-frequency data. However, to avoid inconvenient restrictions on the parameter space, we propose to parametrize the distribution in terms of the overdispersion parameter, i.e.\ the excessive variance. The dynamic overdispersion then follows score-driven model, developed by \cite{Creal2013} and \cite{Harvey2013}. The estimated diurnal pattern of squared price changes and their relation to trade durations are further plugged into this dynamics. The used relation to trade durations simultaneously captures adjustment of variance to time unit and the residual dependency on trade durations, which were modeled separately by \cite{Engle2000}. The proposed joint modeling removes the problems with zero trade durations, which can be quite frequent in high-frequency data. The proposed model belongs to the class of observation-driven models and can be estimated by the maximum likelihood method, which makes it suitable even for large datasets.

In an empirical study, we focus on the IBM stock (just as, e.g., \citealp{Engle1998, Engle2000}) from March to July, 2022. However, we also report results for 6 other stocks traded on the NYSE and NASDAQ exchanges. We estimate intraday models with various specifications for each of the 105 trading days in our dataset. For the IBM stock, the average number of observations in a day is 63\,673. We show that the proposed model is a good fit and all its components are justifiable. As the models are estimated on a daily basis, we proceed to analyze their forecasting performance on the days following the days used for the estimation. We also demonstrate how the results can be used as an alternative to daily realized measures of volatility such as the realized kernel of \cite{Barndorff-Nielsen2008}. Finally, we find that the relation between price volatility and trade durations is the same as described by \cite{Engle2000}, eventhough the magnitude of high-frequency data has increased considerabely since then.

The structure of the paper is outlined as follows. The proposed methodology is detailed in Section \ref{sec:meth}. Specifically, Section \ref{sec:methAdj} presents the preliminary steps of temporal adjustment, Section \ref{sec:methSkellam} describes the zero-inflated Skellam distribution with the overdispersion parameterization, and Section \ref{sec:methDynamics} features the proposed dynamic model with notes on its estimation. The empirical study of the IBM stock is presented in Section \ref{sec:emp}. Specifically, Section \ref{sec:empData} describes the analyzed dataset, Section \ref{sec:empDuration} highlights empirical properties of trade durations, Section \ref{sec:empPrice} focuses on empirical properties of price changes, Section \ref{sec:empIntraday} reports the in-sample performance of the proposed model, Section \ref{sec:empAlt} compares the model to some alternatives, Section \ref{sec:empForecast} conducts a forecasting study and addresses aggregation of simultaneous transactions, and Section \ref{sec:empDaily} demonstrates the use of the model in the estimation of daily volatility. The paper is concluded in Section \ref{sec:con}.

\begin{table}
\centering
\caption{An overview of selected high-frequency time series models and their features---using ultra-high-frequency data with irregularly spaced observations (Irreg), accounting for simultaneous transactions with zero trade durations (Simul), accounting for discrete prices or price changes (Discrete), accounting for market microstructure noise (Noise), joint modeling of volatility and trade durations (Duration), joint modeling of volatility and trade volume (Volume), and multivariate modeling (Multi).}
\label{tab:literature}
\begin{tabular}{lccccccc}
\toprule
Paper & Irreg & Simul & Discrete & Noise & Duration & Volume & Multi \\
\midrule
\cite{Ghysels1998}    & \cmark & \xmark & \xmark & \xmark & \cmark & \xmark & \xmark \\
\cite{Engle2000}      & \cmark & \xmark & \xmark & \xmark & \cmark & \xmark & \xmark \\
\cite{Grammig2002}    & \cmark & \xmark & \xmark & \xmark & \cmark & \xmark & \xmark \\
\cite{Manganelli2005} & \cmark & \xmark & \xmark & \xmark & \cmark & \cmark & \xmark \\
\cite{Russell2005}    & \cmark & \xmark & \cmark & \xmark & \cmark & \xmark & \xmark \\
\cite{Liu2012}        & \cmark & \xmark & \xmark & \cmark & \cmark & \xmark & \xmark \\
\cite{Huptas2016}     & \cmark & \xmark & \xmark & \xmark & \cmark & \xmark & \xmark \\
\cite{Koopman2017a}   & \xmark & \xmark & \cmark & \xmark & \xmark & \xmark & \xmark \\
\cite{Koopman2018}    & \xmark & \xmark & \cmark & \xmark & \xmark & \xmark & \cmark \\
\cite{Buccheri2021}   & \xmark & \xmark & \xmark & \cmark & \xmark & \xmark & \cmark \\
\cite{Catania2022}    & \xmark & \xmark & \cmark & \xmark & \xmark & \xmark & \cmark \\
\cite{Holy2022b}      & \cmark & \xmark & \cmark & \xmark & \xmark & \xmark & \xmark \\
\textbf{This study}   & \cmark & \cmark & \cmark & \cmark & \xmark & \xmark & \xmark \\
\bottomrule
\end{tabular}
\end{table}

\section{Methodology}
\label{sec:meth}

\subsection{Nonparametric Temporal Adjustment}
\label{sec:methAdj}

In our analysis, we examine transactional data. For each transaction indexed by $i = 0, \ldots, n$, we denote the time of the transaction as $t_i$, the price as $p_i$, and the traded volume as $v_i$. We particularly focus on two derived variables: trade durations, $d_i = t_i - t_{i-1}$ for $i = 1, \ldots, n$, and price changes, $y_i = 100 \cdot (p_i - p_{i-1})$ for $i = 1, \ldots, n$. The precision of $p_i$ is limited to two decimal places, resulting in integer values for price changes. Note that unlike log returns, price changes should be interpreted relative to the price level. We treat trade durations as continuous\footnote{Although in our particular dataset presented in Section \ref{sec:emp}, the precision of $t_i$ is only to three decimal places.}.

Empirical evidence widely acknowledges that trade durations exhibit a distinct intraday pattern, characterized by longer durations observed at noon, while shorter durations are more prevalent near the opening and closing of the market. Moreover, high-frequency price changes typically demonstrate an expected value very close to zero, while their variance exhibits a notable intraday pattern, characterized by lower volatility observed during the middle of a day, while higher volatility occurs near the opening and closing of the market. Additionally, there exists a discernible relation between the variance of price changes and trade durations; specifically, longer trade duration are correlated with higher variance (see, e.g., \citealp{Engle2000}).

We start our analysis by estimating the diurnal patterns of trade durations and squared prices changes, along with estimating the relation between squared price changes and trade durations. In Section \ref{sec:methDynamics}, we then model the dynamics of $y_i$ adjusted for the diurnal pattern and its relation to trade durations. Given the assumption of zero expected value for price changes, the observed squared price changes serve as a proxy for variance. Note that when the expected value is zero, the average of squared price changes is a consistent estimator for both the variance and the overdispersion parameter, which is introduced later in Section \ref{sec:methSkellam}. Anticipating nonlinearity, we employ the cubic smoothing spline method (see, e.g., \citealp[Section 5.4]{Hastie2008}). The chosen nonparametric method, however, is not essential to our approach and alternatives can be used as well.

Now, let us discuss diurnal adjustment in more detail. First, we estimate the intraday pattern of trade durations. We assume that the intraday pattern of trade durations is the same on all days but the level of trade durations can differ. For this reason, we standardize trade durations as $\bar{d}_i = d_i / \frac{1}{n} \sum_{i=1}^n d_i$ on each day. Using the complete dataset, we then estimate the dependence of $\bar{d}_i$ on $t_i$ by the cubic smoothing spline method. We obtain the fitted function $\hat{f}_{\mathrm{dur}}(t_i)$ and adjust trade durations as $\tilde{d}_i = \bar{d}_i / \hat{f}_{\mathrm{dur}}(t_i)$.

Next, we estimate the intraday pattern of squared price changes. As in the case of trade durations, we assume that the intraday pattern of squared prices changes is the same on all days but the level of squared prices changes can differ. We standardize squared price changes as $\bar{y}^2_i = y^2_i / \frac{1}{n} \sum_{i=1}^n y^2_i$ on each day and then estimate the dependence of $\bar{y}^2_i$ on $t_i$ by the cubic smoothing spline method using the complete dataset. We obtain the fitted function $\hat{f}_{\mathrm{var}}(t_i)$ and adjust squared price changes as $\tilde{y}^2_i = \bar{y}^2_i / \hat{f}_{\mathrm{var}}(t_i)$.

Finally, we estimate the relation between squared price changes and trade durations, i.e.\ dependence of $\tilde{y}^2_i$ on $\tilde{d}_i$. Again, we use the cubic smoothing spline method and obtain the fitted function $\hat{f}_{\mathrm{rel}}(\tilde{d}_i)$.

\subsection{Zero-Inflated Skellam Distribution}
\label{sec:methSkellam}

The probability theory and statistics literature does not offer many distributions defined on integer support (without the nonnegativity or positivity constraint). The most used representative is the Skellam distribution of \cite{Skellam1946}, which is the distribution of the difference between two independent variables following the Poisson distribution with rates $\lambda_1$ and $\lambda_2$ respectively. Regarding dynamic models, it can be used when a time series of counts is nonstationary, but its first difference is stationary---a typical feature of high-frequency prices.

The Skellam distribution is often parametrized in terms of mean $\mu = \lambda_1 - \lambda_2$ and variance $\sigma^2 = \lambda_1 + \lambda_2$ rather than rates $\lambda_1$ and $\lambda_2$ (see, e.g., \citealp{Koopman2017a, Koopman2018, Alomani2018}). However, in this parametrization, it is required that $\sigma^2 > \lvert \mu \rvert$. When $\mu$ is nonzero, this condition can be hard to satisfy in models with dynamic variance (due to the lower bound on variance $\lvert \mu \rvert$) and/or dynamic mean (due to the lower bound on mean $-\sigma^2$ and the upper bound on mean $\sigma^2$). For this reason, we propose an alternative parametrization with overdispersion parameter $\delta = \sigma^2 - \lvert \mu \rvert = \min \{ 2 \lambda_1, 2 \lambda_2 \} > 0$ representing excessive variance.

The standard Skellam distribution with two parameters, however, lacks the necessary flexibility to effectively model high-frequency prices where the majority of values are concentrated around zero. \cite{Koopman2017a} deflate the probability of 0 and inflate probability of 1 and -1 using an additional parameter. On the other hand, \cite{Karlis2006, Karlis2009} and \cite{Catania2022} inflate the probability of 0 and deflate the probabilities of all other values using an additional parameter. As our data exhibit increased occurence of zero values (in comparison to the fitted Skellam distribution), we follow the latter approach and introduce the zero-inflation parameter $\pi$ to the distribution.

Zero inflation is modeled following the approach of \cite{Lambert1992} as:
\begin{equation}
\label{eq:zi}
\begin{aligned}
Y &\sim 0 & & \text{with probability } \pi, \\
Y &\sim \mathrm{Skellam}(\mu, \delta) & & \text{with probability } 1 - \pi.
\end{aligned}
\end{equation}
The probability mass function of the zero-inflated Skellam distribution with the mean-overdispersion parametrization is then given by
\begin{equation}
\label{eq:ziskellamProb}
\mathrm{P} [Y = y \mid \mu, \delta, \pi] = \begin{cases}
\pi + (1 - \pi) \exp( - \lvert \mu \rvert - \delta) I_0 \left( \sqrt{\delta^2 + 2 \lvert \mu \rvert \delta} \right) & \text{ for } y = 0, \\
(1 - \pi) \exp( - \lvert \mu \rvert - \delta) \left( \frac{\lvert \mu \rvert + \mu + \delta}{\lvert \mu \rvert - \mu + \delta} \right)^{\frac{y}{2}} I_y \left( \sqrt{\delta^2 + 2 \lvert \mu \rvert \delta} \right) & \text{ for } y \neq 0, \\
\end{cases}
\end{equation}
where $I_{\cdot}(\cdot)$ is the modified Bessel function of the first kind. The first two moments are given by
\begin{equation}
\label{eq:ziskellamMom}
\mathrm{E}[Y] = (1 - \pi) \mu, \qquad \mathrm{var}[Y] = (1 - \pi) \left( \lvert \mu \rvert + \delta + \pi \mu^2 \right).
\end{equation}

\subsection{Dynamic Model}
\label{sec:methDynamics}

In the dynamic model, we let the mean parameter $\mu$ and the overdispersion parameter $\delta$ be time-varying but keep the zero-inflation parameter $\pi$ static.

Strong negative first order autocorrelation, insignificant autocorrelation of higher order, and decaying negative partial autocorrealtion is typical for ultra-high-frequency price changes or returns and is caused by market microstrucure noise (see, e.g., \citealp{Ait-Sahalia2005, Hansen2006}). It can be effectively captured by MA(1) process. Another typical feature of high-frequency data is zero mean of price changes or returns in long term (see, e.g., \citealp{Koopman2017a}). We therefore model dynamics of the mean parameter as MA(1) process with zero intercept,
\begin{equation}
\label{eq:dynMean}
\mu_i = \theta \left( y_{i-1} - \mu_{i-1} \right),
\end{equation}
where $\theta$ is the moving average parameter.

For the dynamics of the overdispersion parameter, we adopt a GARCH-like structure and include the temporal adjustments presented in Section \ref{sec:methAdj}. To avoid any restrictions on the parameter space, we model the logarithm of the overdispersion parameter, which is in line with the multiplicative form of the temporal adjustments. Similarly to \cite{Koopman2018}, we let the overdispersion parameter be driven by lagged conditional score, i.e.\ the gradient of the log-likelihood, of the Skellam distribution. Our model therefore belongs to the class of score-driven models (see \citealp{Creal2013, Harvey2013})\footnote{Besides \cite{Koopman2018}, score-driven model based on the Skellam distribution was also used by \cite{Koopman2019} in an application to football results.}. All put together, the dynamics of the overdispersion parameter is given by
\begin{equation}
\label{eq:dynDisp}
\ln \left( \delta_i \right) = \omega + \ln \left( \hat{f}_{\mathrm{var}}(t_i) \right) + \ln \left( \hat{f}_{\mathrm{rel}}(\tilde{d}_i) \right) + \varepsilon_i, \qquad \varepsilon_i = \varphi \varepsilon_{i-1} + \alpha \nabla_{\ln(\delta)} \left( y_{i-1}; \mu_{i-1}, \delta_{i-1}, \pi \right),
\end{equation}
where $\omega$ is the intercept, $\varphi$ is the autoregressive parameter, $\alpha$ is the score parameter, and $\nabla_{\ln(\delta)} (\cdot)$ is the score given by
\begin{equation}
\begin{aligned}
\nabla_{\ln(\delta)} (y; \mu, \delta, \pi) &= \frac{\partial \ln \mathrm{P} [Y = y \mid \mu, \delta, \pi]}{\partial \ln(\delta)} \\
&= \begin{cases}
\frac{\delta (\pi - 1) \left( \sqrt{\delta^2 + 2 \lvert \mu \rvert \delta} I_0 \left( \sqrt{\delta^2 + 2 \lvert \mu \rvert \delta} \right) - (\lvert \mu \rvert + \delta) I_1 \left( \sqrt{\delta^2 + 2 \lvert \mu \rvert \delta} \right) \right)}{\sqrt{\delta^2 + 2 \lvert \mu \rvert \delta} \left( (1 - \pi) I_0 \left( \sqrt{\delta^2 + 2 \lvert \mu \rvert \delta} \right) + \pi \exp(\lvert \mu \rvert + \delta) \right)} & \text{ for } y = 0, \\
\frac{\delta^2 + \lvert \mu \rvert \delta}{2 \sqrt{\delta^2 + 2 \lvert \mu \rvert \delta}} \frac{ I_{y-1} \left( \sqrt{\delta^2 + 2 \lvert \mu \rvert \delta} \right) + I_{y+1} \left( \sqrt{\delta^2 + 2 \lvert \mu \rvert \delta} \right) }{ I_y \left( \sqrt{\delta^2 + 2 \lvert \mu \rvert \delta} \right) } - \frac{\mu y}{\delta + 2 \lvert \mu \rvert} - \delta & \text{ for } y \neq 0. \\
\end{cases}
\end{aligned}
\end{equation}
Although the formula for the score is quite complex in the case of the Skellam distribution, its interpretation is clear: it measures the discrepancy between an observation $y$ and the probability distribution implied by parameters $\mu$, $\delta$, and $\pi$. In the dynamics, it acts as a correction term quantifying the direction and magnitude by which $\ln(\delta)$ should change in order to improve the fit of the distribution.

Note that \eqref{eq:dynMean} can be rewritten as 
\begin{equation}
\label{eq:dynDispAdj}
\ln \left( \frac{\delta_i}{\hat{f}_{\mathrm{var}}(t_i) \hat{f}_{\mathrm{rel}}(\tilde{d}_i)} \right) = \omega + \varepsilon_i, \qquad \varepsilon_i = \varphi \varepsilon_{i-1} + \alpha \nabla_{\ln(\delta)} \left( y_{i-1}; \mu_{i-1}, \delta_{i-1}, \pi \right).
\end{equation}
This formulation illustrates that dynamics are specified for the logarithm of the overdispersion parameter, which is initially adjusted for the intraday pattern by $\hat{f}_{\mathrm{var}}(t_i)$ and then further adjusted for the current trade duration by $\hat{f}_{\mathrm{rel}}(\tilde{d}_i)$. A common approach in financial econometrics is a two-step procedure, where returns are initially adjusted for temporal effects, and subsequently, the analysis is conducted on the adjusted values (see, e.g., \citealp{Engle2000}). However, this method is not appropriate for discrete price differences, as the adjustment would yield non-integer values. Therefore, we incorporate the adjustment terms directly into the dynamics equation. Nonetheless, $\hat{f}_{\mathrm{var}}(t_i)$ and $\hat{f}_{\mathrm{rel}}(\tilde{d}_i)$ are estimated separately in advance to facilitate computation.

The timing of transactions is crucial for modeling volatility. In our dynamics, $\delta_i$ depends on $t_i$ and $d_i$, which are both available only after transaction $i$ occurs. This limits the use of the proposed model in forecasting. The formulation \eqref{eq:dynDispAdj} express dynamics in terms of the adjusted overdispersion parameter, which requires only information up to transaction $i-1$. Therefore, our model can effectively forecast values of the adjusted overdispersion parameter. However, for forecasting the unadjusted overdispersion parameter, an additional model forecasting the timing of transactions, such as the ACD model, would be required.

There are five parameters in the model to be estimated---$\theta$, $\omega$, $\varphi$, $\alpha$, and $\pi$. The model is observation-driven and we find the parameters by maximizing the log-likelihood,
\begin{equation}
\ell \left(\theta, \omega, \varphi, \alpha, \pi \mid y_1, \ldots, y_n \right) = \frac{1}{n} \sum_{i=1}^n \ln \mathrm{P} [Y_i = y_i \mid \mu_i, \delta_i, \pi],
\end{equation}
where $\mu_i$ and $\delta_i$ are given by \eqref{eq:dynMean} and \eqref{eq:dynDisp} respectively. As $\mu_i$ and $\varepsilon_i$ are defined recursively, it is needed to set their initial values. We set them to their long-term average, i.e.\ $\mu_0 = \varepsilon_0 = 0$\footnote{The particular choice for the initialization is not, however, that important as their impact quickly fades out and is overall negligible in the tens of thousands or even hundreds of thousands of observations we have.}. We numerically find the optimal values of the parameters using the Nelder--Mead algorithm. It is, however, possible to use any general-purpose algorithm solving nonlinear optimization problems.

Deriving asymptotic properties of the maximum likelihood estimates is beyond the scope of the paper. We refer to \cite{Alzaid2010} for the theoretical results on static case of the Skellam distribution and \cite{Blasques2018,Blasques2022} for the results on score-driven models in general. Tailoring these results to our specific model is, however, not straightforward.

\section{Empirical Study}
\label{sec:emp}

\subsection{Analyzed Data Sample}
\label{sec:empData}

As \cite{Engle1998}, \cite{Engle2000}, and many other papers, we focus our analysis on the IBM stock traded on the New York Stock Exchange (NYSE). The stock is included in the Dow Jones Industrial Average (DJIA), S\&P 100, and S\&P 500 indices. We use tick-by-tick transaction data from March to July, 2022---a total of 105 trading days. The source of the data is Refinitiv Eikon\footnote{Formerly operated by Thomson Reuters.}. Furthermore, we report results for the CAT, MA, and, MCD stocks traded on NYSE and the CSCO, EA, and INTC stocks traded on NASDAQ in Appendix \ref{app:further}.

We preform standard data cleaning steps, as described e.g.\ in \cite{Barndorff-Nielsen2009}. Namely, we remove observations outside the standard trading hours 9:30--16:00 EST, remove observations in the first 5 minutes after the opening (we further discuss this in Section \ref{sec:empPrice}), remove observations without recorded price, remove outliers (when price exceeds 10 mean absolute deviations from a rolling centred median of 50 observations), and round prices to the nearest cent.

After data cleaning, we get the total of 6\,685\,657 transactions over 105 trading days for the IBM stock, which corresponds to 2.721 transactions per second. The two busiest days are July 19 with 258\,217 transactions and April 20 with 184\,250 transactions. Both these days follow announcements of quarterly results on July 18 and April 19 respectively. The quietest day is March 28 with just 35\,333 transactions. The median value is 56\,894 transactions per trading day.

The subsequent analysis is performed using R. The temporal adjustment is performed by the \verb"smooth.spline()" function from the \verb"stats" package \citep{RCoreTeam2022}. The dynamic model is estimated by the \verb"gas()" function from the \verb"gasmodel" package \citep{Holy2022} with a one-line modification\footnote{The score for $\mu$ in the zero-inflated Skellam distribution is replaced by $y-\mu$ to mimic the moving average process.}.

\subsection{Trade Durations}
\label{sec:empDuration}

We start the empirical study by a brief look at trade durations. The data are recorded with a time precision of one millisecond and we report trade durations in seconds (with precision to three decimal places). It should be noted that the data are recorded sequentially---the order of transactions with the same timestamp matters. We can think of observations as having a true unique timestamp of high precision, which is then lost, retaining only the timestamp of low precision and the original ordering. 

\begin{figure}
\centering
\includegraphics[width=15cm]{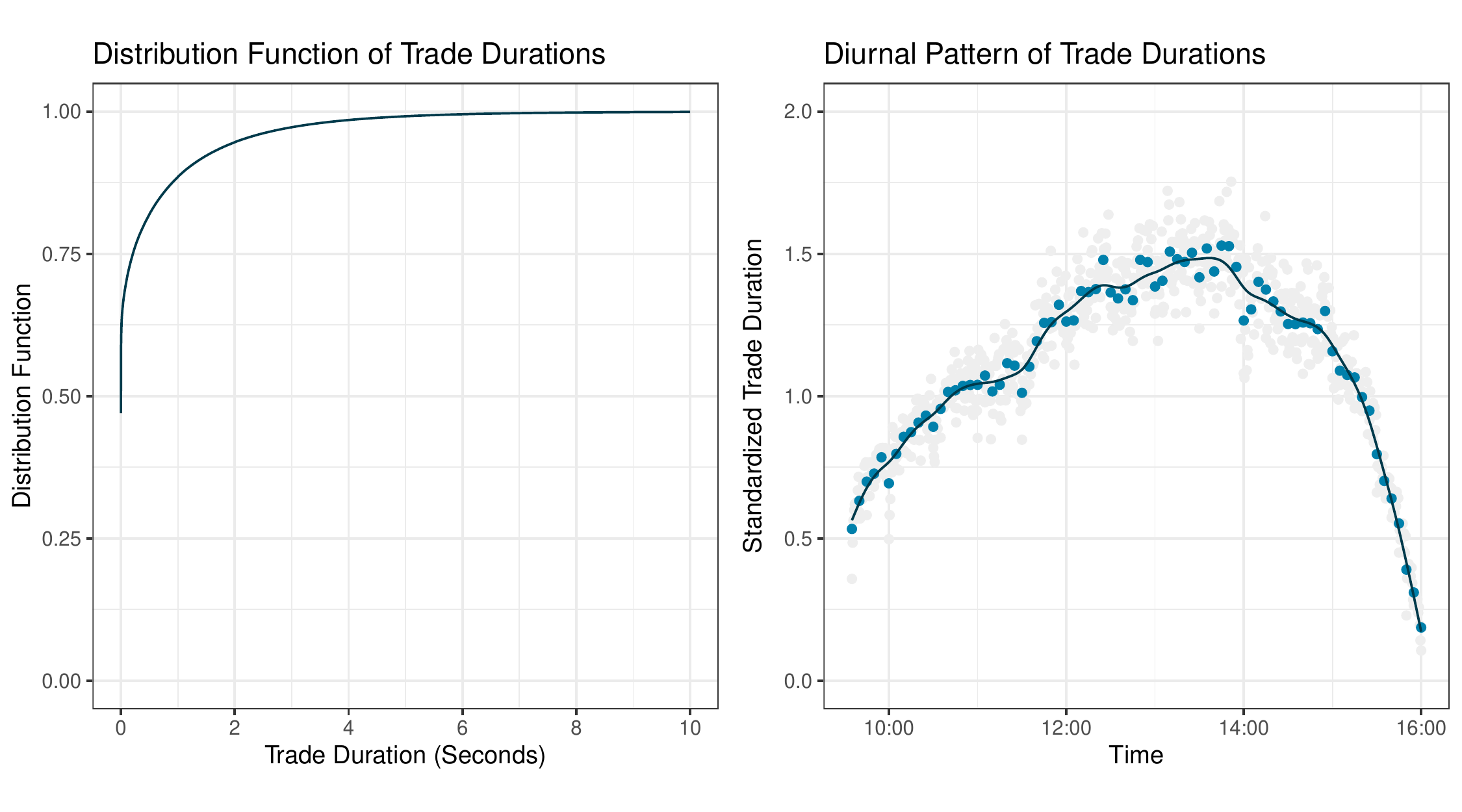}
\caption{Left: The empirical distribution function of trade durations. Right: The average trade durations in 5 minute (blue dots) and 30 second (grey dots) intraday intervals with a smoothed curve $\hat{f}_{\mathrm{dur}}(t_i)$. The results are for the IBM stock.}
\label{fig:duration}
\end{figure}

The left plot of Figure \ref{fig:duration} shows the empirical distribution of trade durations. Most transactions occur in close succession---47 percent of trade durations are equal to zero, 62 percent are lower than one centisecond, 70 percent are lower than one decisecond, and 88 percent are lower than one second. Thus, aggregating simultaneous transactions would almost halve the number of observations. Using a similar dataset for the IBM stock, \cite{Blasques2022a} found that 95 percent of zero trade durations are caused by split transactions while 5 percent are unrelated transactions. We decide to keep simultaneous transactions in our dataset, but further address this issue in Section \ref{sec:empForecast}.

The right plot of Figure \ref{fig:duration} shows diurnal pattern of trade durations---a typical hill shape. The market is most active after opening and before closing while after noon there is a quiet period. This is consistent with the duration literature.

\subsection{Price Changes}
\label{sec:empPrice}

\begin{figure}
\centering
\includegraphics[width=15cm]{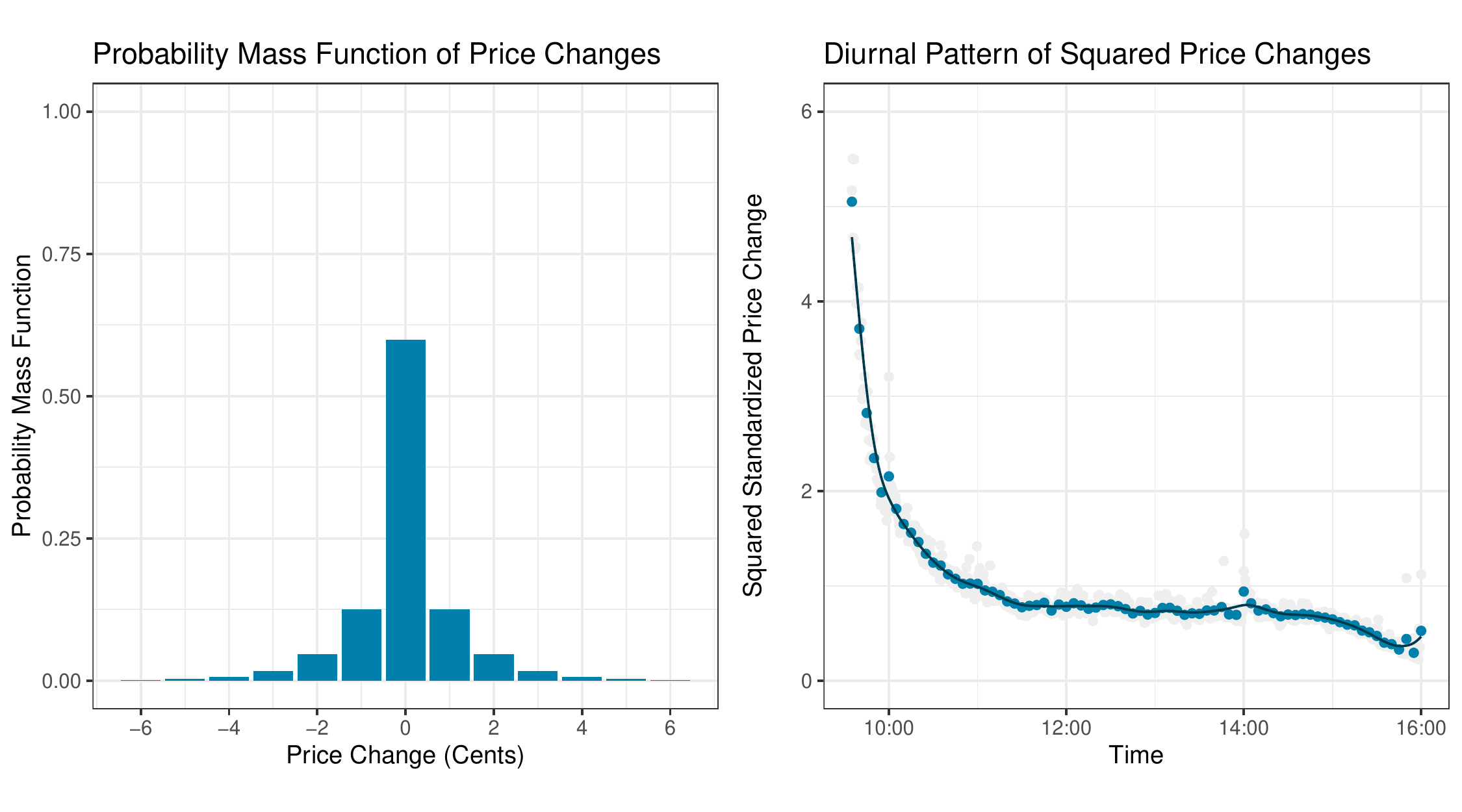}
\caption{Left: The empirical probability mass function of price changes. Right: The average squared price changes in 5 minute (blue dots) and 30 second (grey dots) intraday intervals with a smoothed curve $\hat{f}_{\mathrm{var}}(t_i)$. The results are for the IBM stock.}
\label{fig:return}
\end{figure}

Next, we move on to empirical properties of price changes. The left plot of Figure \ref{fig:return} shows the empirical probability mass function of price changes. The price changes at ultra-high-frequency are quite low---60 percent of price changes are zero and 99 percent of price changes are between -3 and 3 cents. Only 0.002 percent of price changes lie outside the $[-20,20]$ interval. The most extreme price changes are -66 and 68 cents\footnote{The occurrence of these high price changes that are not marked as outliers is associated with overall high volatility in their neighborhood, and consequently, these observations pass the test for outliers. Many other high price changes were, however, labeled as outliers and removed during data cleaning.}. The average price of the IBM stock is 133 USD in the analyzed period. The most extreme price changes are therefore about 0.5 percent of the price.

The right plot of Figure \ref{fig:return} shows diurnal pattern of squared price changes. There is extreme volatility after the opening, which quickly declines. As smoothing splines have trouble capturing this steep decrease, we remove the first 5 minutes from data, i.e.\ we focus only on 9:35--16:00 EST. Right before the closing, volatility slightly increases. There is also a slight increase around 14:00 associated with news relevant to the IBM stock\footnote{In the case of the IBM stock, the increase is not that major. In the case of other stocks, however, this could be much larger jump (or multiple jumps at various times), which smoothing splines could fail to capture; see Appendix \ref{app:further}.}.

There is strong serial correlation present in both price changes and squared price changes. The autocorrelation of price changes is -0.352 for the first order and very close to zero for higher orders. The partial autocorrelation, on the other hand, decreases gradually. This is a well-known stylized fact of high-frequency data studied, e.g., by \cite{Ait-Sahalia2005} and \cite{Hansen2006}. The autocorrelation of squared price changes is 0.403 for the first order and gradually decreases for higher orders. The partial autocorrelation also decreases gradually. This suggests MA(1) dynamics for the mean process and richer dynamics for the volatility process.

\begin{figure}
\centering
\includegraphics[width=15cm]{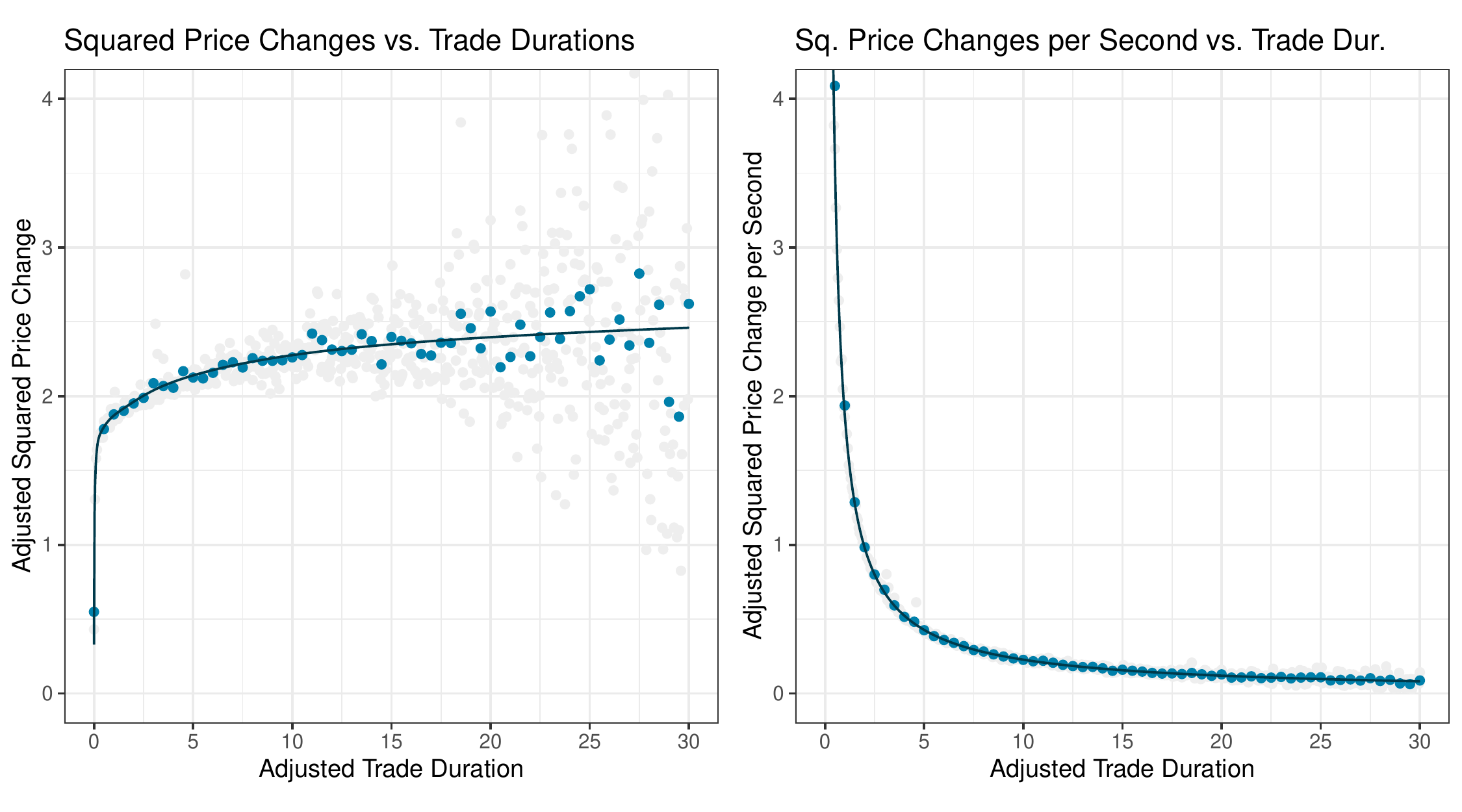}
\caption{Left: The average diurnally adjusted squared price changes in half second (blue dots) and 50 millisecond (grey dots) intervals of diurnally adjusted trade durations with a smoothed curve $\hat{f}_{\mathrm{rel}}(\tilde{d}_i)$. Right: The average diurnally adjusted squared price changes per second in half second (blue dots) and 50 millisecond (grey dots) intervals of diurnally adjusted trade durations with a smoothed curve $\hat{f}_{\mathrm{rel}}(\tilde{d}_i) / \tilde{d}_i$. The results are for the IBM stock.}
\label{fig:relation}
\end{figure}

Under the assumption of a random walk and independent times of transactions, the variance of price changes should linearly increase with trade duration. However, as visualized in the left plot of Figure \ref{fig:relation}, we observe that the increase is actually slower than linear. This is further emphasized in the right plot of Figure \ref{fig:relation}, which shows that price variance per second (squared price changes divided by trade duration) is not constant but decreases with trade duration. This is in line with \cite{Engle2000}, who estimated a positive linear dependence of variance per time unit on the inverse of trade duration. \cite{Engle2000} attributes this behavior to the theory of \cite{Easley1992}, simply summarized as ``no trade means no news,'' which suggests that longer durations are associated with no relevant news and thus stable prices with lower volatility per time unit.

In our model, unlike \cite{Engle2000}, we do not compute the variance (or the overdispersion parameter) per time unit, which would require positivity of trade durations. A counterpart to the volatility model of \cite{Engle2000}\footnote{Equation (39) in \cite{Engle2000}.} within our framework\footnote{The notation, the overdispersion parameter instead of the variance, the multiplicative form via the logarithm, the separate component for the dynamics.} would be:
\begin{equation}
\label{eq:dynEngle}
\ln \left( \frac{\delta_i}{\tilde{d}_i} \right) = \omega + \ln \left( \hat{f}_{\mathrm{var}}(t_i) \right) + \gamma \ln \left( \frac{1}{\tilde{d}_i} \right) + \varepsilon_i, \qquad \varepsilon_i = \varphi \varepsilon_{i-1} + \alpha \nabla_{\ln(\delta)} \left( y_{i-1}; \mu_{i-1}, \delta_{i-1}, \pi \right).
\end{equation}
Here, we would be dividing by zero twice---when calculating the overdispersion parameter per second and when inverting trade durations. Note that for the purposes of the right plot of Figure \ref{fig:relation}, we add 0.001 to the values of trade durations. Of course, this is a completely arbitrary transformation, which has a significant impact on behavior near zero (which is cropped in the right plot of Figure \ref{fig:relation}). Instead, we directly estimate the relation between the squared price changes and trade durations as $\hat{f}_{\mathrm{rel}}(\tilde{d}_i)$ and thus avoid problems with zero values of trade durations.

\subsection{Dynamics of Intraday Price Volatility}
\label{sec:empIntraday}

For each of the 105 trading days in our sample, we estimate 5 models---the proposed model along with its variants which set different parameters to zero. The ``naive model'' is the simplest one, with restrictions set as $\theta = \varphi = \alpha = \pi = 0$, meaning it solely estimates the level $\omega$ of the overdispersion parameter. However, it also incorporates nonparametric temporal adjustments. The ``no inflation model`` imposes $\pi = 0$, the ``static dispersion model'' imposes $\varphi = \alpha = 0$, and the ``static mean model'' imposes $\theta = 0$. The ``proposed model'' is the most general one, allowing for zero inflation with dynamics given by \eqref{eq:dynMean} and \eqref{eq:dynDisp}. 

There are several reasons why we estimate models separately for each day rather than all days together. Nowadays, UHF data consists of a vast number of observations, and due to computational restrictions, it is often necessary to utilize only a portion of the available data in a single model. Our dataset, spanning 105 trading days, consists of 6\,685\,657 transactions. A trading day is a natural time period characterized by uninterrupted trading, with bounds defined by the opening and closing of the market. Furthermore, the characteristics of data can change over time, necessitating the re-estimation of the model. We study the change of the estimated models on day-by-day basis in Section \ref{sec:empForecast}.

In Table \ref{tab:coef}, we report the minimum, maximum, and median values from the 105 sets of estimated parameters, as we estimate the models on a daily basis. Note that, we do not report p-values as all parameters are significant due to huge numbers of observations (with the exception of $\pi$ for a single day, as further mentioned below). In Table \ref{tab:fit}, we assess fit of the 5 models using the average log-likelihood and residual autocorrelation tests in the form of R$^2$ statistic. Note that the number of parameters (in any of our models) is negligible compared to the number of observations. For this reason, we do not report AIC or BIC.

As discussed in Section \ref{sec:empPrice}, the autocorrelation and partial autocorrelation functions of price changes suggest MA(1) structure for the mean process. Indeed, restricting $\theta$ to zero causes considerable decrease in log-likelihood as evident between the model with static mean and the proposed model. The autocorrelation in residuals also significantly increases. As expected, the estimated $\theta$ is negative for all trading days. The comparison of log-likelihood and autocorrelation in squared residuals between the model with static overdispersion parameter and the proposed model reveals that volatility should not be treated as constant. In each model allowing for zero inflation, $\pi$ is positive for all days except one, July 28\footnote{However, other stocks may exhibit different behaviors, and zero inflation may not be necessary; see Appendix \ref{app:further}.}. This suggests that there is an increased occurence of zero price changes in general and the underlying distribution should accomodate this. Among the three components studied in this section---dynamic mean, dynamic volatility, and zero inflation---setting parameter $\pi$ to zero decreases the log-likelihood the least, but still distinctly.

Overall, the proposed model performs the best in terms of the log-likelihood among our 5 candidates. The proposed specification for the mean and overdispersion processes also overwhelmingly reduces residual autocorrelation in price changes and squared price changes. Due to huge number of observations, however, it is difficult to obtain statistical significance of no autocorrelation. The associated Ljung--Box test rejects no autocorrelation in residuals of the proposed model for all days and lags at 0.01 significance level. The associated ARCH-LM test suggests no autocorrelation in squared residuals of the proposed model for 55 percent of days for lag 1 but only 4 percent for lag 100 at 0.01 significance level. Nevertheless, the R$^2$ static is very low in all cases and the model captures mean and volatility dynamics quite well.

All 5 models utilize adjustments for the time of the day $\hat{f}_{\mathrm{var}}(t_i)$ and the current trade duration $\hat{f}_{\mathrm{rel}}(\tilde{d}_i)$. If we remove both types of temporal adjustments from the proposed model, we obtain the average log-likelihood of -1.256. If we remove only $\hat{f}_{\mathrm{rel}}(\tilde{d}_i)$ and keep $\hat{f}_{\mathrm{var}}(t_i)$, we obtain the average log-likelihood of -1.255. From this, we can see that adjusting the overdispersion parameter for the time of the day increases the fit only marginally, and the diurnal pattern visualized in Figure \ref{fig:return} can be effectively captured by our score-driven dynamics. Adjusting for the current trade duration, however, is a crucial step in our approach that significantly increases the fit of the model.

\begin{table}
\centering
\caption{The minimum, median, and maximum values of the estimated parameters of the models estimated on daily basis. The results are for the IBM stock.}
\label{tab:coef}
\footnotesize
\begin{tabular}{llrrrrr}
\toprule
& & \multicolumn{5}{c}{Model} \\
\cmidrule(l{3pt}r{3pt}){3-7}
Coef. & Trans. & \multicolumn{1}{c}{Naive} & \multicolumn{1}{c}{No Infl.} & \multicolumn{1}{c}{Static Disp.} & \multicolumn{1}{c}{Static Mean} & \multicolumn{1}{c}{Proposed} \\
\midrule
          & Min &  & -0.454 & -0.575 &  & -0.536 \\ 
$\theta$  & Med &  & -0.305 & -0.386 &  & -0.354 \\ 
          & Max &  & -0.223 & -0.304 &  & -0.268 \\ \\
          & Min & -0.379 & -0.730 & -0.796 & -0.400 & -0.730 \\ 
$\omega$  & Med & 0.200 & -0.163 & -0.052 & 0.275 & -0.024 \\ 
          & Max & 0.676 & 0.382 & 0.656 & 0.860 & 0.621 \\ \\
          & Min &  & 0.937 &  & 0.644 & 0.939 \\ 
$\varphi$ & Med &  & 0.974 &  & 0.834 & 0.975 \\ 
          & Max &  & 0.995 &  & 0.984 & 0.995 \\ \\
          & Min &  & 0.078 &  & 0.104 & 0.091 \\
$\alpha$  & Med &  & 0.168 &  & 0.498 & 0.191 \\ 
          & Max &  & 0.246 &  & 0.681 & 0.276 \\ \\
          & Min &  &  & 0.000 & 0.000 & 0.000 \\ 
$\pi$     & Med &  &  & 0.143 & 0.119 & 0.134 \\ 
          & Max &  &  & 0.260 & 0.235 & 0.231 \\
\bottomrule
\end{tabular}
\end{table}

\begin{table}
\centering
\caption{The R$^2$ statistics of the residuals and the squared residuals regressed on their lagged values with the average log-likelihood of an observation for the models estimated on daily basis. The results are for the IBM stock.}
\label{tab:fit}
\footnotesize
\begin{tabular}{llrrrrr}
\toprule
& & \multicolumn{5}{c}{Model} \\
\cmidrule(l{3pt}r{3pt}){3-7}
Statistic & Lag & \multicolumn{1}{c}{Naive} & \multicolumn{1}{c}{No Infl.} & \multicolumn{1}{c}{Static Disp.} & \multicolumn{1}{c}{Static Mean} & \multicolumn{1}{c}{Proposed} \\
\midrule
               &   1 & 0.118 & 0.004 & 0.002 & 0.077 & 0.003 \\ 
AR R$^2$       &  10 & 0.151 & 0.008 & 0.005 & 0.097 & 0.006 \\ 
               & 100 & 0.154 & 0.011 & 0.007 & 0.098 & 0.009 \\ \\
               &   1 & 0.104 & 0.000 & 0.001 & 0.005 & 0.000 \\ 
ARCH R$^2$     &  10 & 0.150 & 0.004 & 0.030 & 0.007 & 0.003 \\ 
               & 100 & 0.181 & 0.008 & 0.050 & 0.016 & 0.006 \\ \\
Log-Lik.       &     & -1.264 & -1.181 & -1.188 & -1.212 & -1.174 \\ 
\bottomrule
\end{tabular}
\end{table}

\subsection{Comparison to Alternative Models}
\label{sec:empAlt}

We compare the proposed model with 5 additional alternatives. We report their log-likelihood and residual autocorrelation statistics in Table \ref{tab:alt}, with a similar structure to Table \ref{tab:fit}.

First, let us focus on the parametrization of the Skellam distribution. We compare the mean-variance parametrization, used e.g.\ by \cite{Koopman2017a, Koopman2018} and \cite{Alomani2018}, with the proposed mean-overdispersion parametrization. The mean-variance parametrization uses the variance parameter $\sigma^2 > |\mu|$ instead of the overdispersion parameter $\delta > 0$. Both parameters are related by $\sigma^2 = \delta + |\mu|$. As in the proposed model, zero-inflation parameter $\pi$ is assumed to be static and the dynamics of $\mu$ are given by \eqref{eq:dynMean}, while the dynamics of $\sigma^2$ are given by
\begin{equation}
\ln \left( \sigma^2_i \right) = \omega + \ln \left( \hat{f}_{\mathrm{var}}(t_i) \right) + \ln \left( \hat{f}_{\mathrm{rel}}(\tilde{d}_i) \right) + \varepsilon_i, \qquad \varepsilon_i = \varphi \varepsilon_{i-1} + \alpha \nabla_{\ln(\sigma^2)} \left( y_{i-1}; \mu_{i-1}, \sigma^2_{i-1}, \pi \right).
\end{equation}
When the mean is not dynamic and is set to zero, both parametrizations are equivalent. When the mean is dynamic, however, Tables \ref{tab:fit} and \ref{tab:alt} show that the mean-overdispersion parametrization is superior in terms of the fitted log-likelihood. The problem, of course, lies in bounds on parameter space imposed by the mean-variance parametrization. In our implementation, we find coefficient values that offer the best fit while ensuring the positivity of the variance for all observations in the sample. However, if the model is used beyond the sample, it may result in zero or negative variance. This prohibits forecasting and renders the parameterization generally unsuitable. The estimated $\theta$ ranges from -0.181 to -0.057, which is considerably lower than that of the proposed model, which ranges from -0.536 to -0.268. This suppression of $\theta$ is caused by the in-sample lower bound on the variance process. In the mean-overdispersion parametrization, there is no such restriction and the mean process is able to reach its full potential. Similarly, there is a difference in estimated values of $\alpha$ and $\varphi$ between the mean-variance and mean-overdispersion parametrizations. The proposed model has higher persistence in comparison to the mean-variance model. Again, this can be atributed to the in-sample lower bound on the variance process in the mean-variance parametrization.

Next, let us further investigate other types of dynamics. As suggested by the autocorelation and partial autocorrelation plots, the proposed model uses the MA(1) structure for the mean. We now specify the mean using the score-driven dynamics, similarly to the time-varying overdispersion parameter, i.e.
\begin{equation}
\label{eq:altMean}
\mu_i = \kappa + \rho \mu_{i-1} + \theta \nabla_{\mu} \left( y_{i-1}; \mu_{i-1}, \delta_{i-1}, \pi \right).
\end{equation}
Compared to \eqref{eq:altMean}, there are two additional parameters $\kappa$ and $\rho$ for the intercept and the autoregressive term, respectively. Additionally, the score is used instead of the lagged error $y_{i-1} - \mu_{i-1}$. Table \ref{tab:alt} shows that this model performs quite poorly with the log-likelihood much lower than for the proposed model. The intercept is redundant, while the autoregressive and score terms does not capture well the dynamics. The dependence of price differences on their lagged values lies in the market microstructure noise, which is best captured by including the $y_{i-1} - \mu_{i-1}$ term.

The proposed model assumes the zero-inflation parameter $\pi$ to be static. We now introduce the score-driven dynamics to this parameter, i.e.
\begin{equation}
\label{eq:altInfl}
\mathrm{logit} (\pi_i) = \gamma + \psi \mathrm{logit} (\pi_{i-1}) + \eta \nabla_{\mathrm{logit} (\pi)} \left( y_{i-1}; \mu_{i-1}, \delta_{i-1}, \pi_{i-1} \right).
\end{equation}
This model is more general than the proposed one, but Table \ref{tab:alt} shows that the average log-likelihood increases only marginally, from -1.174 to -1.172. However, this behavior is not consistent among other stocks\footnote{In Appendix \ref{app:further}, some stocks exhibit a significant increase in the log-likelihood while others only a slight increase. Recall that there are some stocks that do not show zero inflation at all for most days.} and more in-depth study of the occurrence of zero values and its potential relation to other factors is needed. We leave this for future research.

Finally, let us compare our proposed discrete model with continuous approach. We estimate a model based on the normal distribution with time-varying mean $\mu_i$ and variance $\sigma^2_i$,
\begin{equation}
\label{eq:altNorm}
\begin{aligned}
\mu_i &= \theta \left( y_{i-1} - \mu_{i-1} \right), \\
\ln \left( \sigma^2_i \right) &= \omega + \ln \left( \hat{f}_{\mathrm{var}}(t_i) \right) + \ln \left( \hat{f}_{\mathrm{rel}}(\tilde{d}_i) \right) + \varepsilon_i, \qquad \varepsilon_i = \varphi \varepsilon_{i-1} + \alpha \nabla_{\ln(\sigma^2)} \left( y_{i-1}; \mu_{i-1}, \sigma^2_{i-1} \right).
\end{aligned}
\end{equation}
This dynamics is equivalent to \eqref{eq:dynMean} and \eqref{eq:dynDisp} in the proposed model. To compare the fit of this continous model with discrete models, we report the ``discretized'' log-likelihood as
\begin{equation}
\begin{aligned}
\label{eq:loglikCont}
\ell \left(\theta, \omega, \varphi, \alpha, \mid y_1, \ldots, y_n \right) &= \frac{1}{n} \sum_{i=1}^n \ln \mathrm{P} [y_i - 0.5 < Y_i \leq y_i + 0.5 \mid \mu_i, \sigma^2_i] \\
 &= \frac{1}{n} \sum_{i=1}^n \ln \left( F_{Y_i} \left( y_i + 0.5 \mid \mu_i, \sigma^2_i \right) - F_{Y_i} \left(y_i - 0.5 \mid \mu_i, \sigma^2_i \right) \right).
\end{aligned}
\end{equation}
Note that the model is still estimated by maximizing the likelihood based on densities. Table \ref{tab:alt} shows that the model based on the normal distribution has a much lower log-likelihood \eqref{eq:loglikCont} of -1.233 than the proposed model and, in fact, any other model based on the Skellam distribution, except for the naive static model without zero inflation.

We also estimate a more general continuous model based on the Student's t-distribution with dynamics \eqref{eq:altNorm} and additional (static) parameter $\nu$ representing the degrees of freedom. Table \ref{tab:alt} reports strikingly low log-likelihood \eqref{eq:loglikCont} of -3.103 for this model. The model based on the normal distribution, a special case of the Student's t-distribution, thus provides much better fit in terms of log-likelihood \eqref{eq:loglikCont}. Of course the log-likehood based on densities, which is actually maximized, is higher for the general model: -1.202 for the normal distribution and 106.150 for the Student's t-distribution. The estimation of Student's t-distribution degenerates to extremely low values of parameters $\omega$ (median of -368.434) and $\nu$ (median of 0.010). These close-to-zero degrees of freedom $\nu$ cause the mean and the variance not to exist\footnote{Without finite first moment, the residuals cannot be properly defined. In this case, Table \ref{tab:alt} reports NA values for the R$^2$ statistics, which are based on the residuals.}. The density is highly concentrated at zero, but the extremely heavy tails ensure that numerically positive density exists for other values. This behavior highlights the unsuitability of employing a continuous distribution.

\begin{table}
\centering
\caption{The R$^2$ statistics of the residuals and the squared residuals regressed on their lagged values with the average log-likelihood of an observation for the alternative models estimated on daily basis. The results are for the IBM stock.}
\label{tab:alt}
\footnotesize
\begin{tabular}{llrrrrr}
\toprule
& & \multicolumn{5}{c}{Model} \\
\cmidrule(l{3pt}r{3pt}){3-7}
Statistic & Lag & \multicolumn{1}{c}{Var. Param.} & \multicolumn{1}{c}{GAS Mean} & \multicolumn{1}{c}{GAS Infl.} & \multicolumn{1}{c}{Normal} & \multicolumn{1}{c}{Student's-t} \\
\midrule
               &   1 & 0.041 & 0.032 & 0.003 & 0.006 & NA \\ 
AR R$^2$       &  10 & 0.055 & 0.040 & 0.007 & 0.011 & NA \\ 
               & 100 & 0.057 & 0.042 & 0.009 & 0.013 & NA \\ \\
               &   1 & 0.003 & 0.002 & 0.000 & 0.001 & NA \\ 
ARCH R$^2$     &  10 & 0.007 & 0.005 & 0.004 & 0.002 & NA \\ 
               & 100 & 0.018 & 0.012 & 0.007 & 0.006 & NA \\ \\             
Log-Lik.       &     & -1.193 & -1.196 & -1.172 & -1.233 & -3.103 \\ 
\bottomrule
\end{tabular}
\end{table}

\subsection{Simultaneous Transactions and Forecasting}
\label{sec:empForecast}

We shift our focus to the out-of-sample performance of the proposed model. We also assess the influence of aggregating simultaneous transactions, as done e.g.\ by \cite{Engle2000}, compared to keeping all transactions in the dataset, as done so far in our approach. Our model is capable of handling both scenarios. \cite{Engle2000} attributed simultaneous transactions to split transactions, which are large trades broken into two or more smaller trades (see, e.g., \citealp{Pacurar2008}). In more recent datasets, however, zero durations do not necessarily correspond to split transactions. It is possible that two unrelated transactions occur at the same time (especially when the precision of the timestamp is low, as in our case). Conversely, it is possible that split transactions are executed with delays, resulting in a positive duration. Ideally, only split transactions would be aggregated, while unrelated transactions would all be retained in the dataset. Neither of the two approaches ensures this. Without additional information from the limit order book, it is not possible to distinguish between split and unrelated transactions.

The design of this section is as follows. Additionaly to estimating the naive and proposed models on the full datasets, we re-estimate the two models on the dataset with aggregated transactions occurring at the exactly same time (within our precision to milliseconds). The aggregated price is the rounded average of prices with the same timestamp weighted by their volume $v_i$. Note that this transformation roughly reduces the size of the dataset by half (the median number of observations in a day goes from  56\,894 to 31\,070). We then evaluate performance of the models estimated on a specific day on the data from the next day. We thus investigate how the estimated models change on a day-by-day basis. We consider both full and aggregated samples for the forecasting period.

Table \ref{tab:fcst} reports the average log-likelihood as well as the mean absolute error (MAE) and the root mean square error (RMSE). For the naive model, the MAE (and RMSE) is the same whether the model was estimated using the full or aggregated data sample, as the expected value is always zero. For the proposed model, the MAE (and RMSE) is only slightly different. The differences in the average log-likelihood are more pronounced. Unsurprisingly, the models trained on the full sample perform better on the full test sample, and the models trained on the aggregated sample perform better on the aggregated test sample.

Both the naive and proposed models exhibit only slightly lower out-of-sample likelihood compared to in-sample, decreasing from -1.264 to -1.268 for the naive model and from -1.174 to -1.175 for the proposed model, when not aggregating simultaneous transactions. This suggests that the estimated coefficients of the models do not change significantly in the short run. Interestingly, the MAE is lower for the naive model, which predicts only zeros, compared to the proposed model, whose mean is influenced by the market microstructure noise. The RMSE penalizes larger errors more heavily and is lower for the proposed model, as expected.

\begin{table}
\centering
\caption{The average log-likelihood, mean absolute error (MAE), and root mean squared error (RMSE) in the test sample for the naive and proposed models estimated on daily basis. The results are for the IBM stock.}
\label{tab:fcst}
\footnotesize
\begin{tabular}{llrrrr}
\toprule
& & \multicolumn{2}{c}{Full Train Sample} & \multicolumn{2}{c}{Agg. Train Sample} \\
\cmidrule(l{3pt}r{3pt}){3-4} \cmidrule(l{3pt}r{3pt}){5-6}
Test Sample & Statistic & \multicolumn{1}{c}{Naive} & \multicolumn{1}{c}{Proposed} & \multicolumn{1}{c}{Naive} & \multicolumn{1}{c}{Proposed} \\
\midrule
           & Log-Lik. & -1.268 & -1.175 & -1.327 & -1.186 \\ 
Full       & MAE      &  0.645 &  0.681 &  0.645 &  0.681 \\ 
           & RMSE     &  1.220 &  1.133 &  1.220 &  1.132 \\ \\
           & Log-Lik. & -1.747 & -1.600 & -1.672 & -1.589 \\ 
Aggregated & MAE      &  0.994 &  1.018 &  0.994 &  1.018 \\ 
           & RMSE     &  1.549 &  1.446 &  1.549 &  1.445 \\     
\bottomrule
\end{tabular}
\end{table}

\subsection{Daily Measures of Price Volatility}
\label{sec:empDaily}

The proposed approach can naturally be used to model intraday dynamics of prices but also to estimate volatility at daily level as a model-based alternative to various nonparametric volatility measures. A standard nonparametric measure of daily volatility is the realized variance---the sum of squared returns. However, this measure is biased by market microstructure noise and generally not recommended to use at ultra-high-frequency (see, e.g., \citealp{Hansen2006}). At lower frequency such as 5 minutes, however, it can be sufficient as the impact of market microstructure noise is reduced (see, e.g., \citealp{Liu2015}). A widely used realized measure that is robust to market microstructure noise is the realized kernel of \cite{Barndorff-Nielsen2008}\footnote{For details on practical use of the realized kernel, see \cite{Barndorff-Nielsen2009}. For the multivariate case, see \cite{Barndorff-Nielsen2011a}. Other noise-robust realized measures such as the multi-scale and pre-averaging estimators are fairly similar as they can all be expressed in a quadratic form (see, e.g., \citealp{Holy2023a}).}.

In this section, we compare the realized variance and the realized kernel based on the modified Tukey--Hanning kernel with realized measures implied by our model. The total variance based on the proposed model is given by
\begin{equation}
TMV = \sum_{i=1}^n (1 - \pi) \left( \lvert \mu_i \rvert + \delta_i + \pi \mu_i^2 \right).
\end{equation}
Following \eqref{eq:dynDispAdj}, we can also measure volatility by the total overdispersion adjusted for temporal effects (both diurnal and duration) given by
\begin{equation}
AMV = \sum_{i=1}^n \frac{\delta_i}{\hat{f}_{\mathrm{var}}(t_i) \hat{f}_{\mathrm{rel}}(\tilde{d}_i)} = \sum_{i=1}^n \exp \left( \omega + \varepsilon_i \right).
\end{equation}
In the latter realized measure, market microstructure noise is filtered by removing the MA(1) component and the effect of trade durations. Note that while the realized variance, realized kernel, and total model volatility are all in the same units of aggregated variance of price changes over the day, the adjusted model volatility is in different units due to the use of the overdispersion parameter and its adjustment for temporal effects.

\begin{figure}
\centering
\includegraphics[width=15cm]{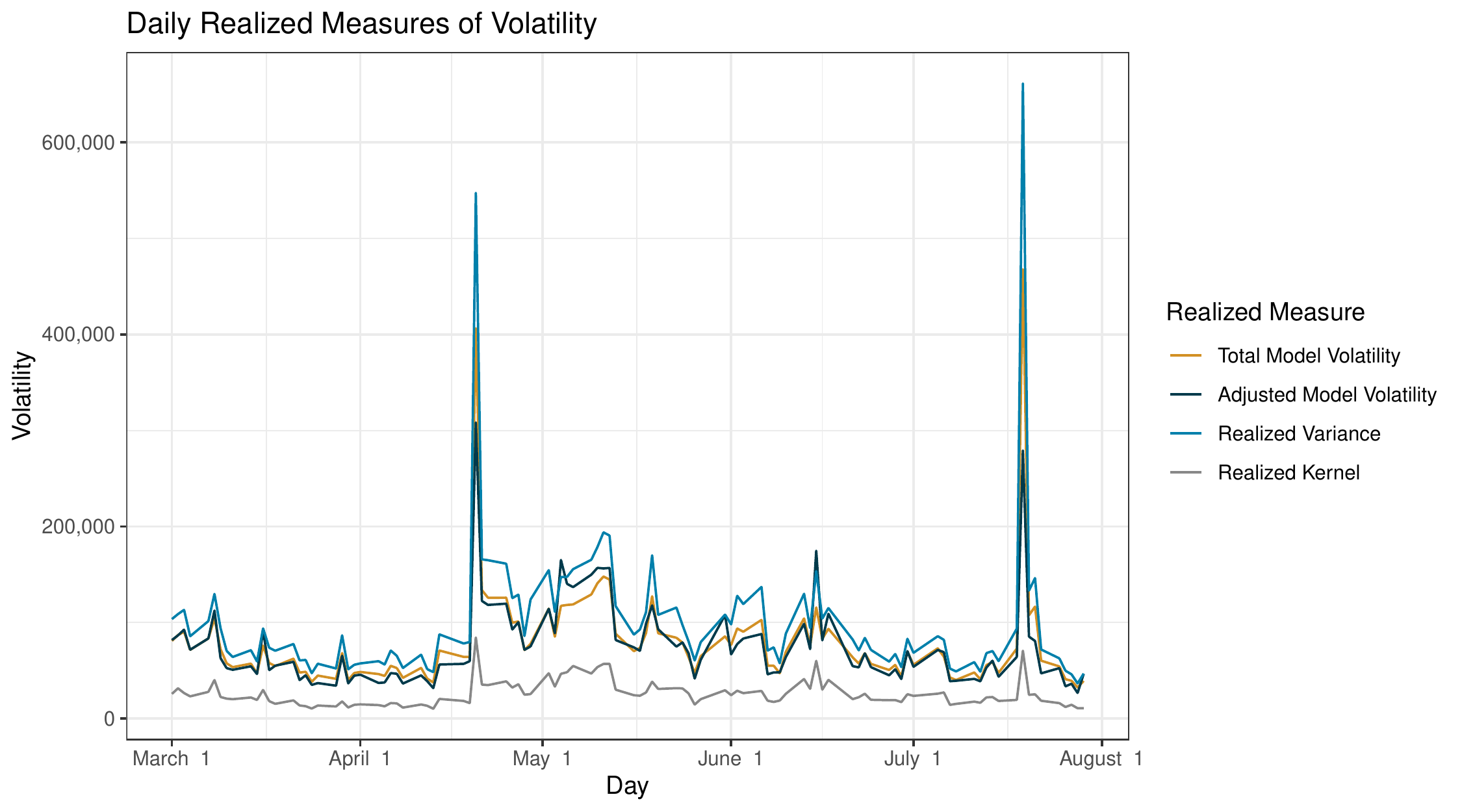}
\caption{The daily values of various volatility realized measures. The results are for the IBM stock.}
\label{fig:realized}
\end{figure}

Figure \ref{fig:realized} shows daily volatility obtained by these measures. The largest variance for all measures is on April 20 (following the announcement of the first quarter results on April 19) and on July 19 (following the announcement of the second quarter results on July 18). We can see that all measures tend to move together but have different scale. This is also supported by a simple correlation analysis. The highest correlations are 0.998 between the total model volatility and the realized variance and 0.961 between the adjusted model volatility and the realized kernel. Other correlations lie between 0.821 and 0.923. We can conclude that the total model volatility is similar to the realized variance as they are both influenced by market microstructure noise. On the other hand, the adjusted model volatility is robust to market microstructure noise, just as the realized kernel. The main benefit of the proposed model-based approach is that we can decompose the variance into individual components according to \eqref{eq:ziskellamMom} and \eqref{eq:dynDispAdj}.

This comparison to fundamentally different approach of realized measures can serve as a sanity check of the proposed model. We can see that the variance given by the proposed model behaves in agreement with realized variance and realized kernel.

\section{Conclusion}
\label{sec:con}

We have proposed a dynamic model for intraday stock prices that takes into account irregularly spaced observations, simultaneous transactions, discreteness of prices, and market microstructure noise. In this model, we have combined two streams of the literature dealing with UHF-GARCH and Skellam-GARCH models, respectively, and further developed them. We have shown that the model finds its use not only in analysis of intraday dynamics but also in estimation of daily volatility.

Suggestions for future research follow Table \ref{tab:literature}. Our model can be extended to include dynamics of trade durations and possibly trade volumes. Another direction lies in multivariate modeling. This is, however, quite challenging due to nonsynchronicity of ultra-high-frequency data. Finally, the model can  be extended to incorporate dynamics for zero changes in prices.

\section*{Acknowledgements}
\label{sec:acknow}

Computational resources were supplied by the project "e-Infrastruktura CZ" (e-INFRA LM2018140) provided within the program Projects of Large Research, Development and Innovations Infrastructures.

\section*{Funding}
\label{sec:fund}

The work on this paper was supported by the Czech Science Foundation under project 23-06139S and the personal and professional development support program of the Faculty of Informatics and Statistics, Prague University of Economics and Business.


\appendix

\section{Evidence from Further Stocks}
\label{app:further}

In this appendix, we report the results for additional stocks: Caterpillar (CAT), traded on NYSE with an average of 2.320 transactions per second; Cisco (CSCO), traded on NASDAQ with an average of 5.738 transactions per second; Electronic Arts (EA), traded on NASDAQ with an average of 1.518 transactions per second; Intel (INTC), traded on NASDAQ with an average of 8.683 transactions per second; Mastercard (MA), traded on NYSE with an average of 2.732 transactions per second; and McDonald's (MCD), traded on NYSE with an average of 2.402 transactions per second.

In general, these results closely resemble those observed for the IBM stock. Nonetheless, there are two distinctions. First, while smoothing splines effectively capture the diurnal patterns of price volatility in the IBM stock, they struggle to account for the impact of news events occurring at regular times. This discrepancy is particularly pronounced when analyzing the INTC stock. Nonetheless, this isn't a significant limitation for our study. Second, zero-inflation is not necessary in most days for the CSCO and INTC stocks, which are the two most frequently traded stocks in our sample. Although zero price changes occur more frequently for these stocks compared to others, a regular Skellam distribution suffices. When the score-driven dynamics is introduced to the zero-inflation parameter, the log-likelihood increases rather significantly for the CAT, EA, MA, and MCD stocks, while it increases only slightly for the CSCO and INTC stocks. In other aspects, the results reinforce the implications drawn from the analysis of the IBM stock.

\begin{figure}
\centering
\includegraphics[width=15cm]{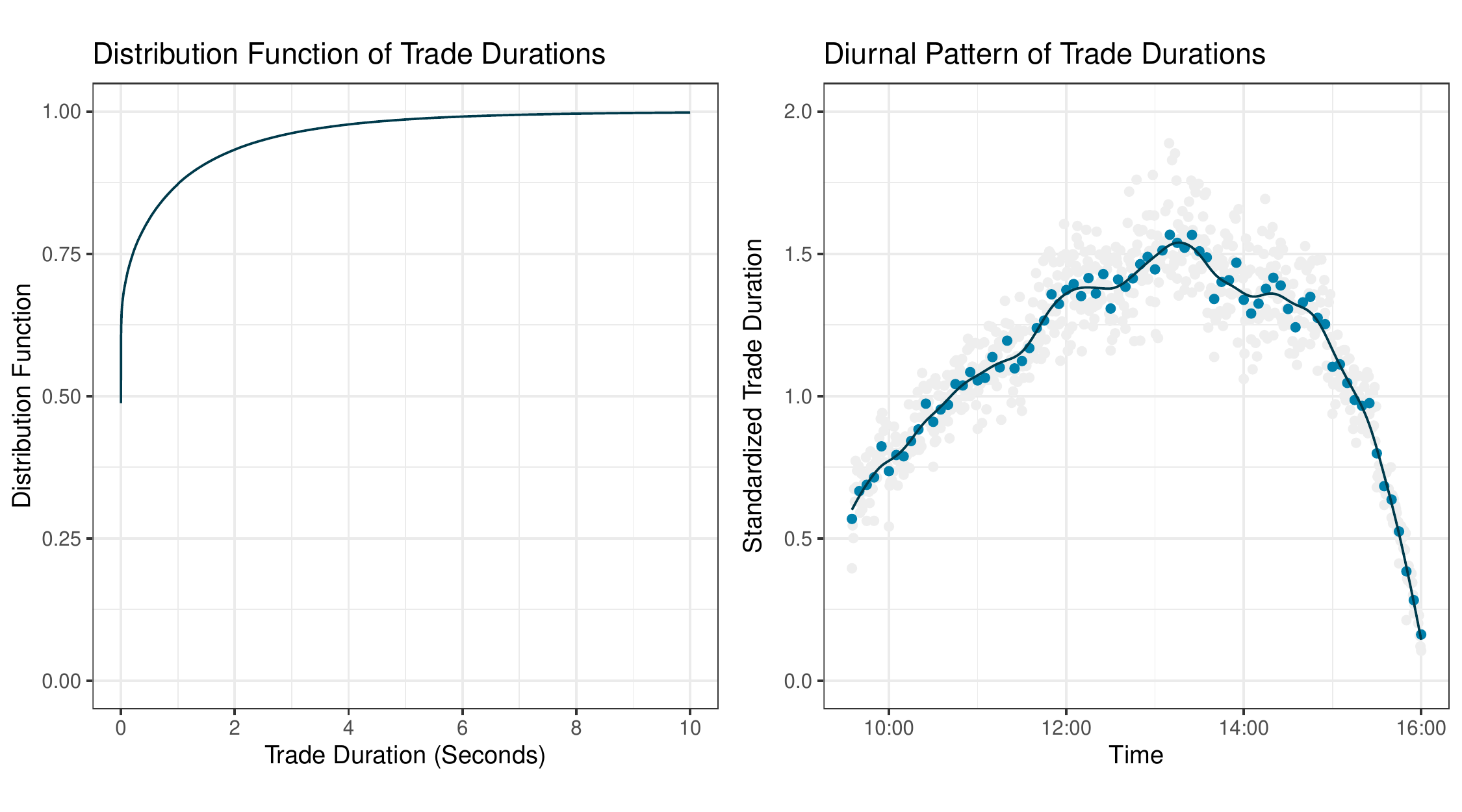}
\caption{Left: The empirical distribution function of trade durations. Right: The average trade durations in 5 minute (blue dots) and 30 second (grey dots) intraday intervals with a smoothed curve $\hat{f}_{\mathrm{dur}}(t_i)$. The results are for the CAT stock.}
\label{fig:durationCAT}
\end{figure}

\begin{figure}
\centering
\includegraphics[width=15cm]{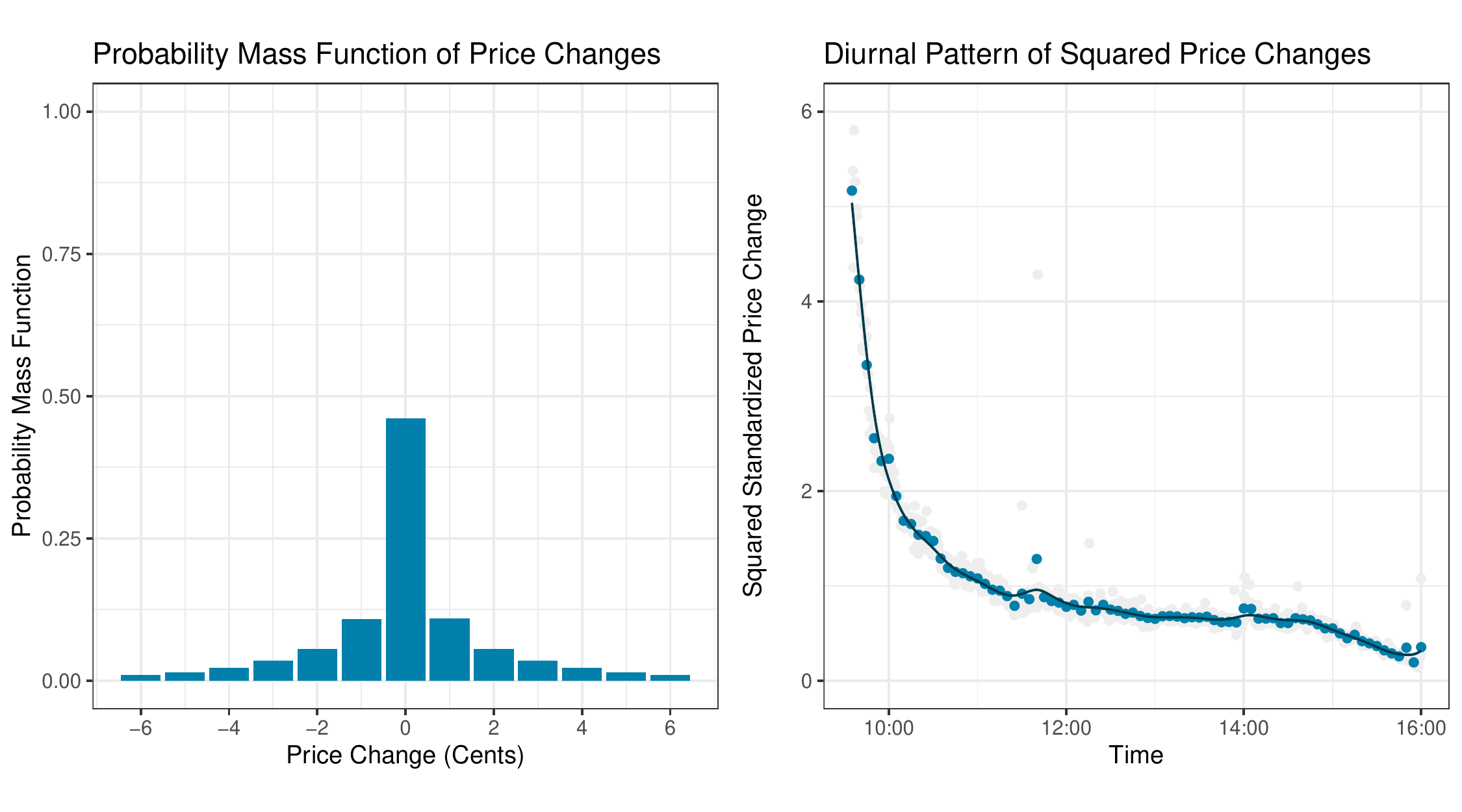}
\caption{Left: The empirical probability mass function of price changes. Right: The average squared price changes in 5 minute (blue dots) and 30 second (grey dots) intraday intervals with a smoothed curve $\hat{f}_{\mathrm{var}}(t_i)$. The results are for the CAT stock.}
\label{fig:returnCAT}
\end{figure}

\begin{figure}
\centering
\includegraphics[width=15cm]{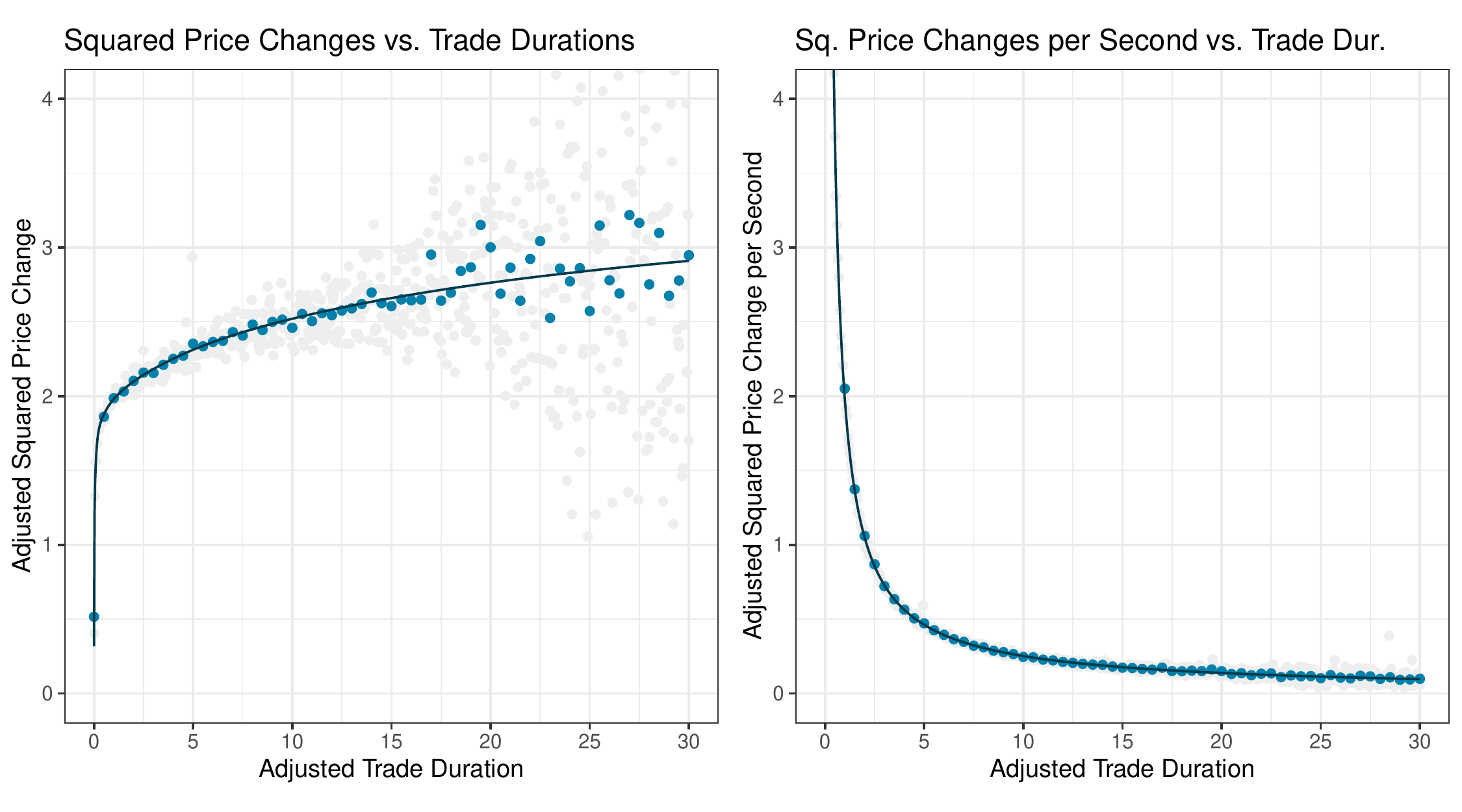}
\caption{Left: The average diurnally adjusted squared price changes in half second (blue dots) and 50 millisecond (grey dots) intervals of diurnally adjusted trade durations with a smoothed curve $\hat{f}_{\mathrm{rel}}(\tilde{d}_i)$. Right: The average diurnally adjusted squared price changes per second in half second (blue dots) and 50 millisecond (grey dots) intervals of diurnally adjusted trade durations with a smoothed curve $\hat{f}_{\mathrm{rel}}(\tilde{d}_i) / \tilde{d}_i$. The results are for the CAT stock.}
\label{fig:relationCAT}
\end{figure}

\begin{table}
\centering
\caption{The minimum, median, and maximum values of the estimated parameters of the models estimated on daily basis. The results are for the CAT stock.}
\label{tab:coefCAT}
\footnotesize
\begin{tabular}{llrrrrr}
\toprule
& & \multicolumn{5}{c}{Model} \\
\cmidrule(l{3pt}r{3pt}){3-7}
Coef. & Trans. & \multicolumn{1}{c}{Naive} & \multicolumn{1}{c}{No Infl.} & \multicolumn{1}{c}{Static Disp.} & \multicolumn{1}{c}{Static Mean} & \multicolumn{1}{c}{Proposed} \\
\midrule
          & Min &  & -0.380 & -0.558 &  & -0.519 \\ 
$\theta$  & Med &  & -0.291 & -0.442 &  & -0.397 \\ 
          & Max &  & -0.223 & -0.359 &  & -0.303 \\ \\
          & Min & 1.122 & 0.780 & 1.107 & 1.318 & 1.069 \\ 
$\omega$  & Med & 1.724 & 1.380 & 1.743 & 1.917 & 1.687 \\ 
          & Max & 2.817 & 2.448 & 2.982 & 3.070 & 2.823 \\ \\
          & Min &  & 0.910 &  & 0.821 & 0.869 \\ 
$\varphi$ & Med &  & 0.954 &  & 0.950 & 0.971 \\ 
          & Max &  & 0.995 &  & 0.996 & 0.997 \\ \\
          & Min &  & 0.026 &  & 0.021 & 0.026 \\ 
$\alpha$  & Med &  & 0.210 &  & 0.199 & 0.203 \\ 
          & Max &  & 0.290 &  & 0.421 & 0.290 \\ \\
          & Min &  &  & 0.215 & 0.177 & 0.191 \\ 
$\pi$     & Med &  &  & 0.269 & 0.236 & 0.242 \\ 
          & Max &  &  & 0.333 & 0.317 & 0.312 \\      
\bottomrule
\end{tabular}
\end{table}

\begin{table}
\centering
\caption{The R$^2$ statistics of the residuals and the squared residuals regressed on their lagged values with the average log-likelihood of an observation for the models estimated on daily basis. The results are for the CAT stock.}
\label{tab:fitCAT}
\footnotesize
\begin{tabular}{llrrrrr}
\toprule
& & \multicolumn{5}{c}{Model} \\
\cmidrule(l{3pt}r{3pt}){3-7}
Statistic & Lag & \multicolumn{1}{c}{Naive} & \multicolumn{1}{c}{No Infl.} & \multicolumn{1}{c}{Static Disp.} & \multicolumn{1}{c}{Static Mean} & \multicolumn{1}{c}{Proposed} \\
\midrule
               &   1 & 0.117 & 0.005 & 0.003 & 0.092 & 0.004 \\ 
AR R$^2$       &  10 & 0.153 & 0.010 & 0.005 & 0.117 & 0.006 \\ 
               & 100 & 0.156 & 0.013 & 0.008 & 0.119 & 0.008 \\ \\
               &   1 & 0.099 & 0.001 & 0.004 & 0.016 & 0.001 \\ 
ARCH R$^2$     &  10 & 0.137 & 0.003 & 0.039 & 0.017 & 0.003 \\ 
               & 100 & 0.173 & 0.008 & 0.065 & 0.022 & 0.006 \\ \\            
Log-Likelihood &     & -2.057 & -1.929 & -1.891 & -1.908 & -1.859 \\ 
\bottomrule
\end{tabular}
\end{table}

\begin{table}
\centering
\caption{The R$^2$ statistics of the residuals and the squared residuals regressed on their lagged values with the average log-likelihood of an observation for the alternative models estimated on daily basis. The results are for the CAT stock.}
\label{tab:altCAT}
\footnotesize
\begin{tabular}{llrrrrr}
\toprule
& & \multicolumn{5}{c}{Model} \\
\cmidrule(l{3pt}r{3pt}){3-7}
Statistic & Lag & \multicolumn{1}{c}{Var. Param.} & \multicolumn{1}{c}{GAS Mean} & \multicolumn{1}{c}{GAS Infl.} & \multicolumn{1}{c}{Normal} & \multicolumn{1}{c}{Student's-t} \\
\midrule
               &   1 & 0.043 & 0.032 & 0.002 & 0.007 & NA \\ 
AR R$^2$       &  10 & 0.058 & 0.038 & 0.005 & 0.014 & NA \\ 
               & 100 & 0.060 & 0.041 & 0.007 & 0.016 & NA \\ \\
               &   1 & 0.008 & 0.005 & 0.000 & 0.002 & NA \\ 
ARCH R$^2$     &  10 & 0.010 & 0.008 & 0.002 & 0.002 & NA \\ 
               & 100 & 0.016 & 0.012 & 0.005 & 0.007 & NA \\ \\
Log-Lik.       &     & -1.881 & -1.884 & -1.848 & -1.991 & -4.310 \\ 
\bottomrule
\end{tabular}
\end{table}

\begin{table}
\centering
\caption{The average log-likelihood, mean absolute error (MAE), and root mean squared error (RMSE) in the test sample for the naive and proposed models estimated on daily basis. The results are for the CAT stock.}
\label{tab:fcstCAT}
\footnotesize
\begin{tabular}{llrrrr}
\toprule
& & \multicolumn{2}{c}{Full Train Sample} & \multicolumn{2}{c}{Agg. Train Sample} \\
\cmidrule(l{3pt}r{3pt}){3-4} \cmidrule(l{3pt}r{3pt}){5-6}
Test Sample & Statistic & \multicolumn{1}{c}{Naive} & \multicolumn{1}{c}{Proposed} & \multicolumn{1}{c}{Naive} & \multicolumn{1}{c}{Proposed} \\
\midrule
           & Log-Lik. & -2.073 & -1.863 & -2.157 & -1.894 \\ 
Full       & MAE      &  1.566 &  1.607 &  1.566 &  1.604 \\ 
           & RMSE     &  2.911 &  2.692 &  2.911 &  2.693 \\ \\
           & Log-Lik. & -2.645 & -2.395 & -2.504 & -2.356 \\ 
Aggregated & MAE      &  2.417 &  2.398 &  2.417 &  2.396 \\ 
           & RMSE     &  3.781 &  3.504 &  3.781 &  3.502 \\
\bottomrule
\end{tabular}
\end{table}

\begin{figure}
\centering
\includegraphics[width=15cm]{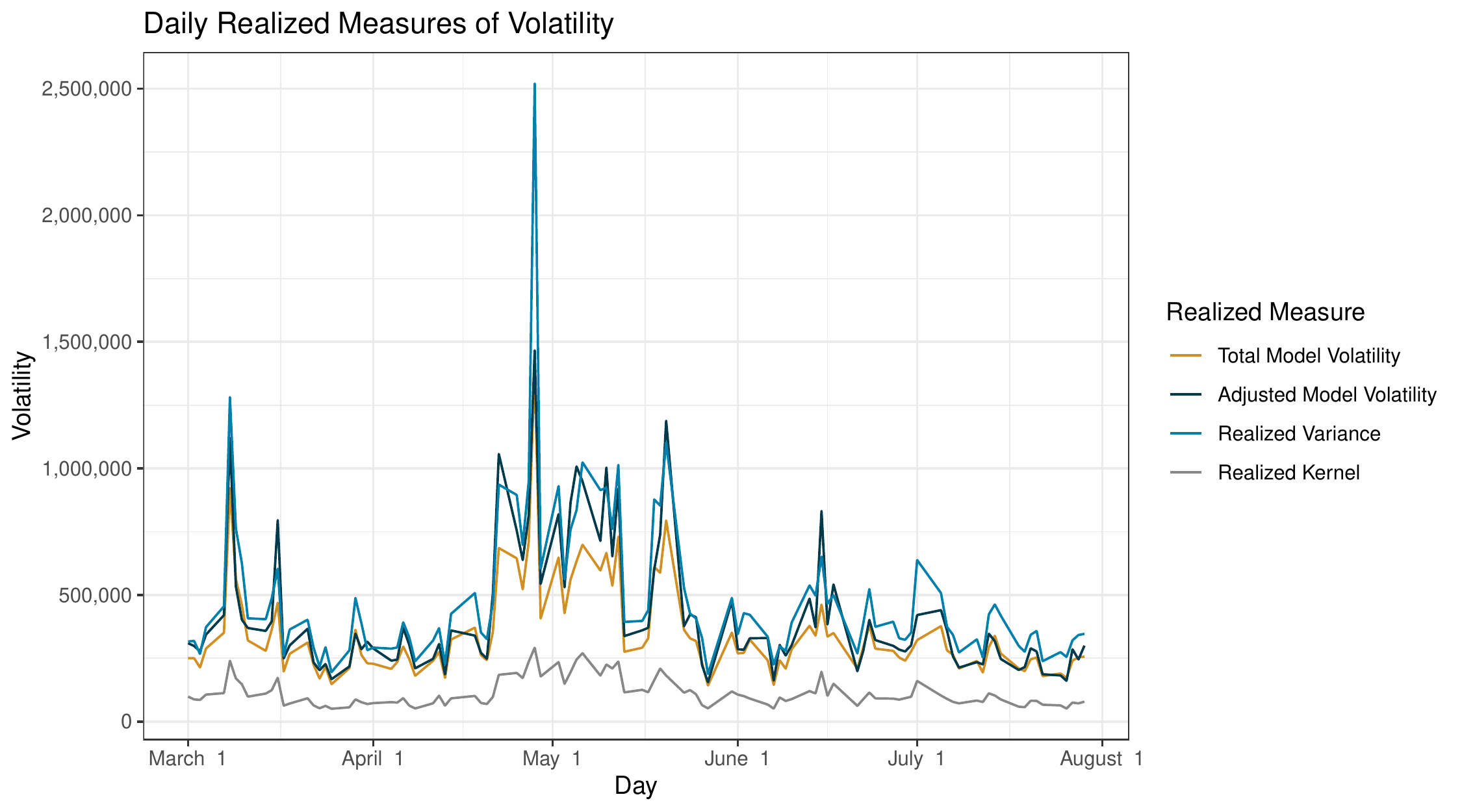}
\caption{The daily values of various volatility realized measures. The results are for the CAT stock.}
\label{fig:realizedCAT}
\end{figure}

\begin{figure}
\centering
\includegraphics[width=15cm]{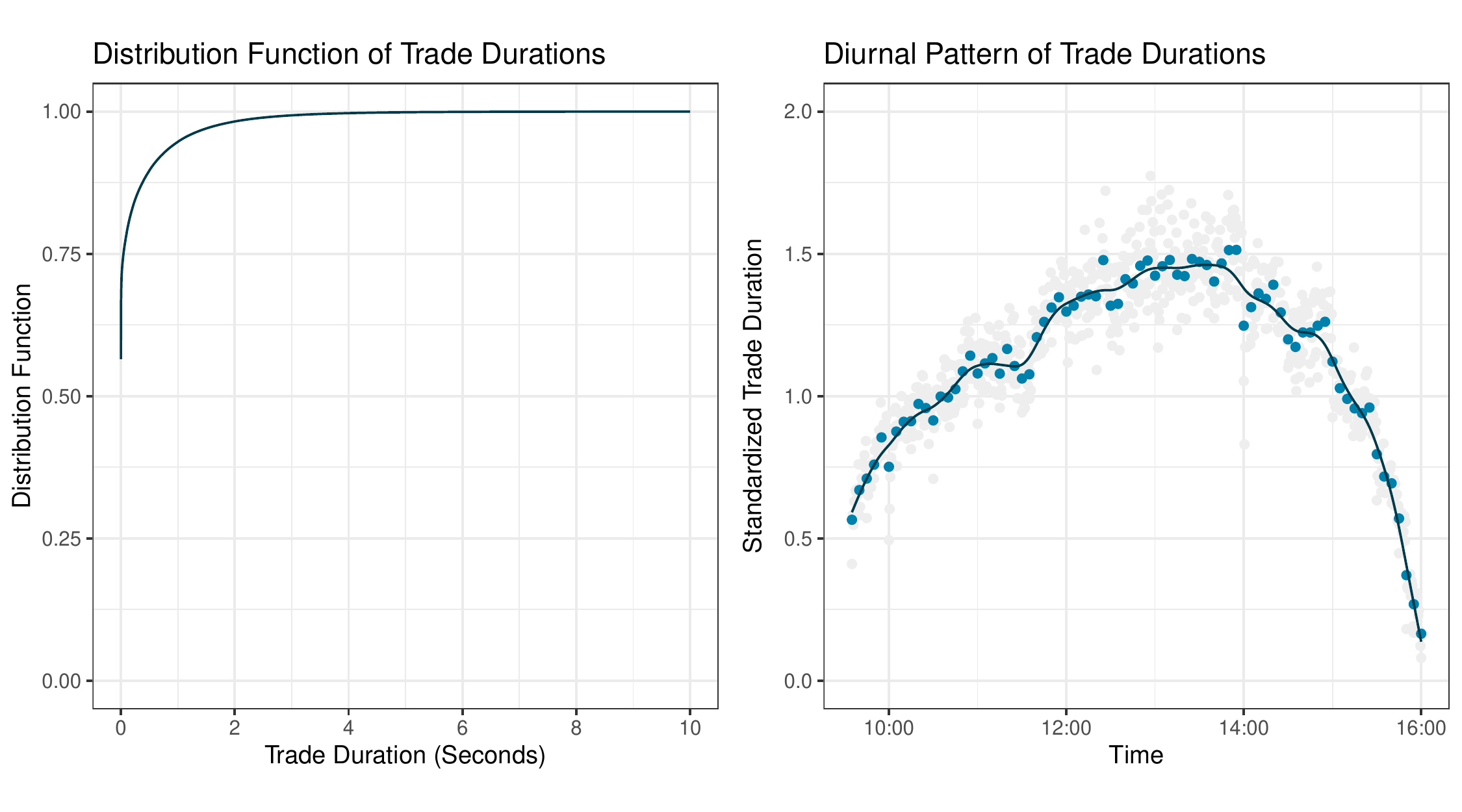}
\caption{Left: The empirical distribution function of trade durations. Right: The average trade durations in 5 minute (blue dots) and 30 second (grey dots) intraday intervals with a smoothed curve $\hat{f}_{\mathrm{dur}}(t_i)$. The results are for the CSCO stock.}
\label{fig:durationCSCO}
\end{figure}

\begin{figure}
\centering
\includegraphics[width=15cm]{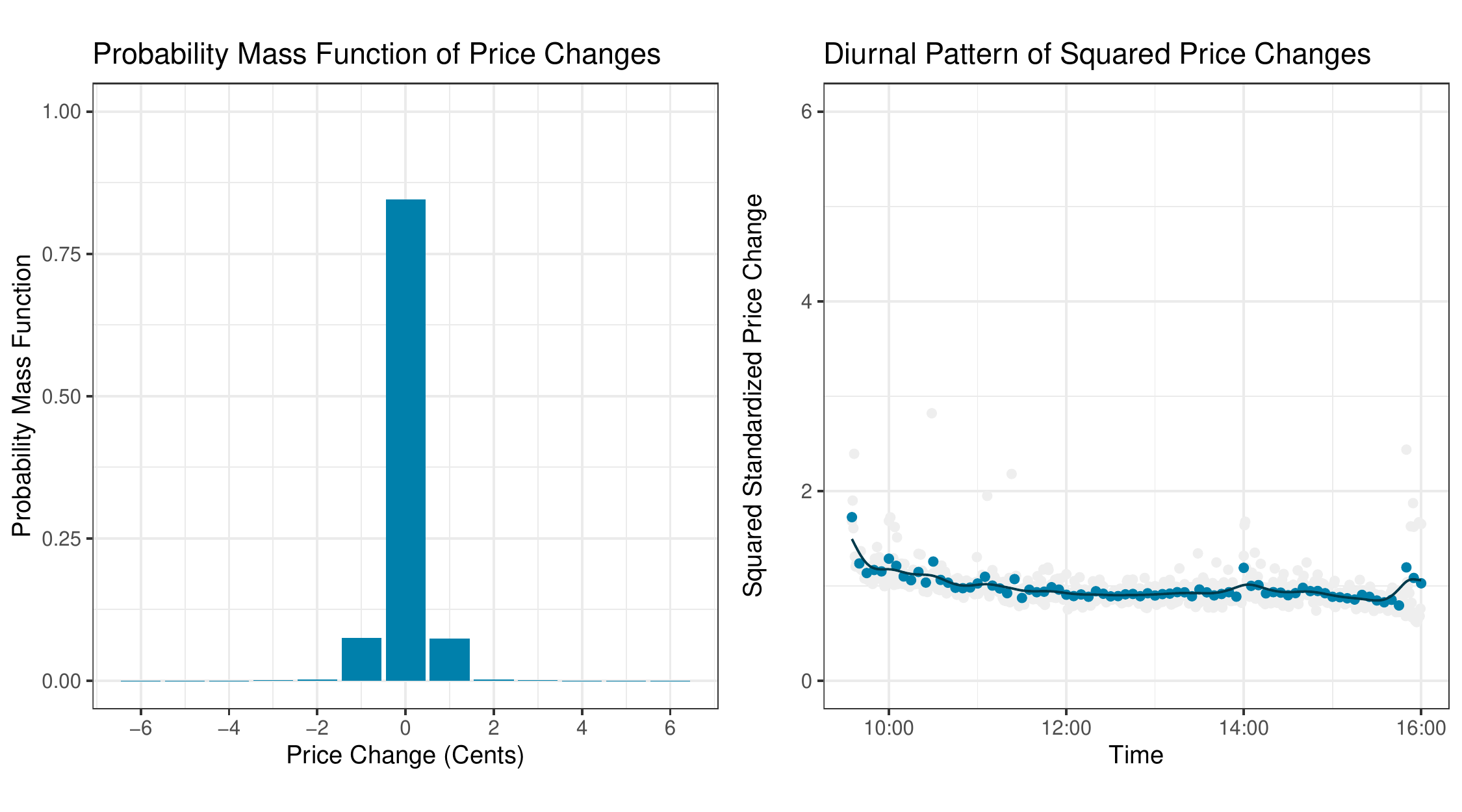}
\caption{Left: The empirical probability mass function of price changes. Right: The average squared price changes in 5 minute (blue dots) and 30 second (grey dots) intraday intervals with a smoothed curve $\hat{f}_{\mathrm{var}}(t_i)$. The results are for the CSCO stock.}
\label{fig:returnCSCO}
\end{figure}

\begin{figure}
\centering
\includegraphics[width=15cm]{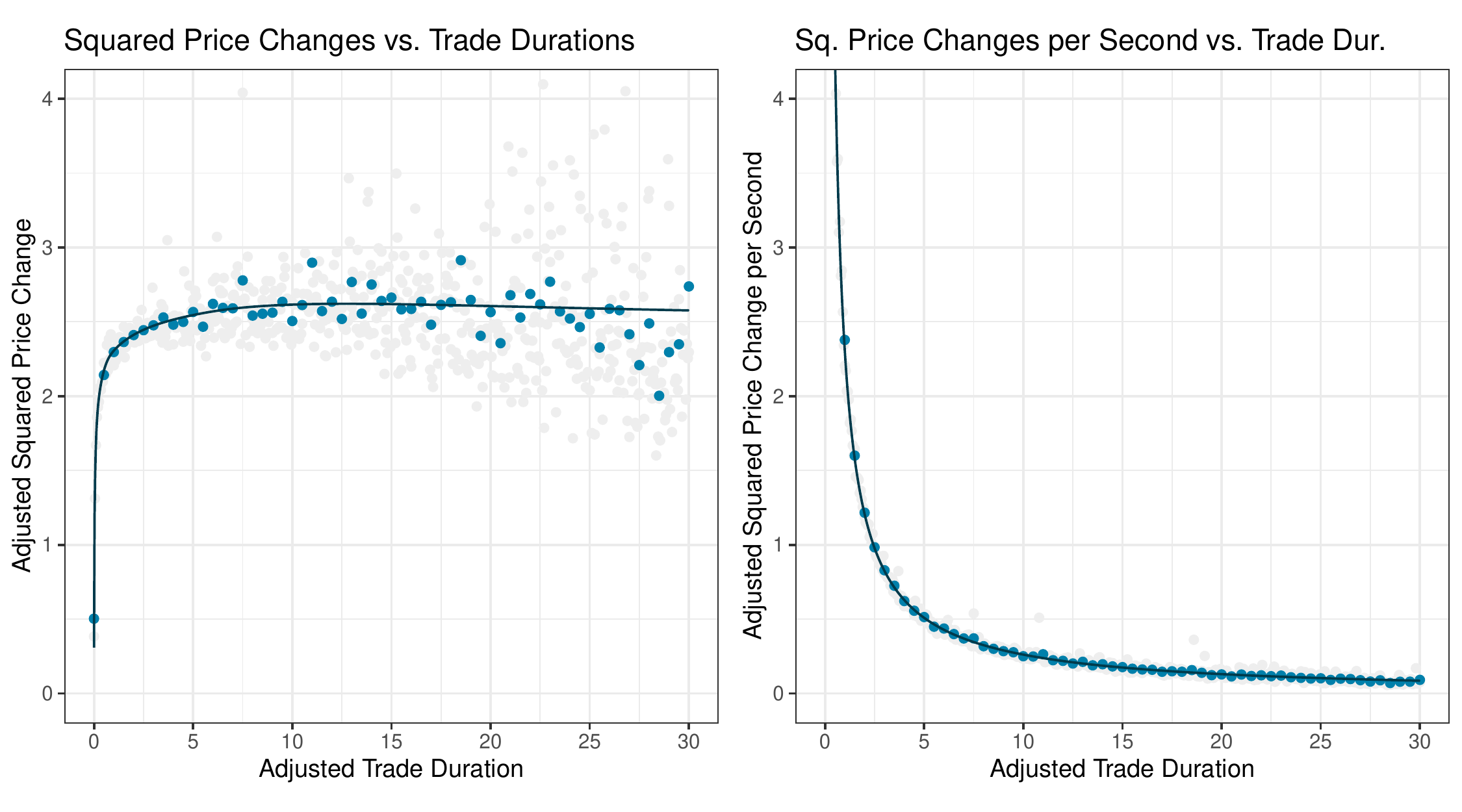}
\caption{Left: The average diurnally adjusted squared price changes in half second (blue dots) and 50 millisecond (grey dots) intervals of diurnally adjusted trade durations with a smoothed curve $\hat{f}_{\mathrm{rel}}(\tilde{d}_i)$. Right: The average diurnally adjusted squared price changes per second in half second (blue dots) and 50 millisecond (grey dots) intervals of diurnally adjusted trade durations with a smoothed curve $\hat{f}_{\mathrm{rel}}(\tilde{d}_i) / \tilde{d}_i$. The results are for the CSCO stock.}
\label{fig:relationCSCO}
\end{figure}

\begin{table}
\centering
\caption{The minimum, median, and maximum values of the estimated parameters of the models estimated on daily basis. The results are for the CSCO stock.}
\label{tab:coefCSCO}
\footnotesize
\begin{tabular}{llrrrrr}
\toprule
& & \multicolumn{5}{c}{Model} \\
\cmidrule(l{3pt}r{3pt}){3-7}
Coef. & Trans. & \multicolumn{1}{c}{Naive} & \multicolumn{1}{c}{No Infl.} & \multicolumn{1}{c}{Static Disp.} & \multicolumn{1}{c}{Static Mean} & \multicolumn{1}{c}{Proposed} \\
\midrule
          & Min &  & -0.572 & -0.619 &  & -0.572 \\ 
$\theta$  & Med &  & -0.467 & -0.481 &  & -0.469 \\ 
          & Max &  & -0.253 & -0.266 &  & -0.253 \\ \\
          & Min & -1.784 & -2.729 & -2.731 & -1.924 & -2.729 \\ 
$\omega$  & Med & -1.659 & -2.325 & -2.349 & -1.736 & -2.325 \\ 
          & Max & -1.466 & -1.896 & -1.910 & 0.002 & -1.896 \\ \\
          & Min &  & 0.956 &  & 0.670 & 0.956 \\ 
$\varphi$ & Med &  & 0.996 &  & 0.843 & 0.996 \\ 
          & Max &  & 1.000 &  & 0.993 & 1.000 \\ \\
          & Min &  & 0.005 &  & 0.113 & 0.005 \\ 
$\alpha$  & Med &  & 0.083 &  & 0.746 & 0.083 \\ 
          & Max &  & 0.279 &  & 1.071 & 0.278 \\ \\
          & Min &  &  & 0.000 & 0.000 & 0.000 \\ 
$\pi$     & Med &  &  & 0.000 & 0.000 & 0.000 \\ 
          & Max &  &  & 0.145 & 0.182 & 0.150 \\           
\bottomrule
\end{tabular}
\end{table}

\begin{table}
\centering
\caption{The R$^2$ statistics of the residuals and the squared residuals regressed on their lagged values with the average log-likelihood of an observation for the models estimated on daily basis. The results are for the CSCO stock.}
\label{tab:fitCSCO}
\footnotesize
\begin{tabular}{llrrrrr}
\toprule
& & \multicolumn{5}{c}{Model} \\
\cmidrule(l{3pt}r{3pt}){3-7}
Statistic & Lag & \multicolumn{1}{c}{Naive} & \multicolumn{1}{c}{No Infl.} & \multicolumn{1}{c}{Static Disp.} & \multicolumn{1}{c}{Static Mean} & \multicolumn{1}{c}{Proposed} \\
\midrule
               &   1 & 0.126 & 0.000 & 0.000 & 0.075 & 0.000 \\ 
AR R$^2$       &  10 & 0.174 & 0.004 & 0.004 & 0.101 & 0.004 \\ 
               & 100 & 0.178 & 0.005 & 0.005 & 0.102 & 0.005 \\ \\
               &   1 & 0.094 & 0.000 & 0.000 & 0.006 & 0.000 \\ 
ARCH R$^2$     &  10 & 0.130 & 0.003 & 0.007 & 0.008 & 0.003 \\ 
               & 100 & 0.158 & 0.007 & 0.021 & 0.015 & 0.007 \\ \\
Log-Likelihood &     & -0.512 & -0.449 & -0.451 & -0.488 & -0.449 \\ 
\bottomrule
\end{tabular}
\end{table}

\begin{table}
\centering
\caption{The R$^2$ statistics of the residuals and the squared residuals regressed on their lagged values with the average log-likelihood of an observation for the alternative models estimated on daily basis. The results are for the CSCO stock.}
\label{tab:altCSCO}
\footnotesize
\begin{tabular}{llrrrrr}
\toprule
& & \multicolumn{5}{c}{Model} \\
\cmidrule(l{3pt}r{3pt}){3-7}
Statistic & Lag & \multicolumn{1}{c}{Var. Param.} & \multicolumn{1}{c}{GAS Mean} & \multicolumn{1}{c}{GAS Infl.} & \multicolumn{1}{c}{Normal} & \multicolumn{1}{c}{Student's-t} \\
\midrule
               &   1 & 0.058 & 0.021 & 0.000 & 0.005 & NA \\ 
AR R$^2$       &  10 & 0.083 & 0.029 & 0.004 & 0.012 & NA \\ 
               & 100 & 0.084 & 0.032 & 0.005 & 0.013 & NA \\ \\
               &   1 & 0.005 & 0.001 & 0.000 & 0.000 & NA \\ 
ARCH R$^2$     &  10 & 0.010 & 0.009 & 0.004 & 0.000 & NA \\ 
               & 100 & 0.019 & 0.013 & 0.008 & 0.003 & NA \\ \\
Log-Lik.       &     & -0.480 & -0.471 & -0.449 & -0.468 & -1.389 \\
\bottomrule
\end{tabular}
\end{table}

\begin{table}
\centering
\caption{The average log-likelihood, mean absolute error (MAE), and root mean squared error (RMSE) in the test sample for the naive and proposed models estimated on daily basis. The results are for the CSCO stock.}
\label{tab:fcstCSCO}
\footnotesize
\begin{tabular}{llrrrr}
\toprule
& & \multicolumn{2}{c}{Full Train Sample} & \multicolumn{2}{c}{Agg. Train Sample} \\
\cmidrule(l{3pt}r{3pt}){3-4} \cmidrule(l{3pt}r{3pt}){5-6}
Test Sample & Statistic & \multicolumn{1}{c}{Naive} & \multicolumn{1}{c}{Proposed} & \multicolumn{1}{c}{Naive} & \multicolumn{1}{c}{Proposed} \\
\midrule
           & Log-Lik. & -0.513 & -0.450 & -0.552 & -0.455 \\ 
Full       & MAE      &  0.165 &  0.195 &  0.165 &  0.195 \\ 
           & RMSE     &  0.454 &  0.414 &  0.454 &  0.414 \\ \\
           & Log-Lik. & -0.925 & -0.799 & -0.870 & -0.795 \\ 
Aggregated & MAE      &  0.318 &  0.363 &  0.318 &  0.363 \\ 
           & RMSE     &  0.624 &  0.571 &  0.624 &  0.571 \\ 
\bottomrule
\end{tabular}
\end{table}

\begin{figure}
\centering
\includegraphics[width=15cm]{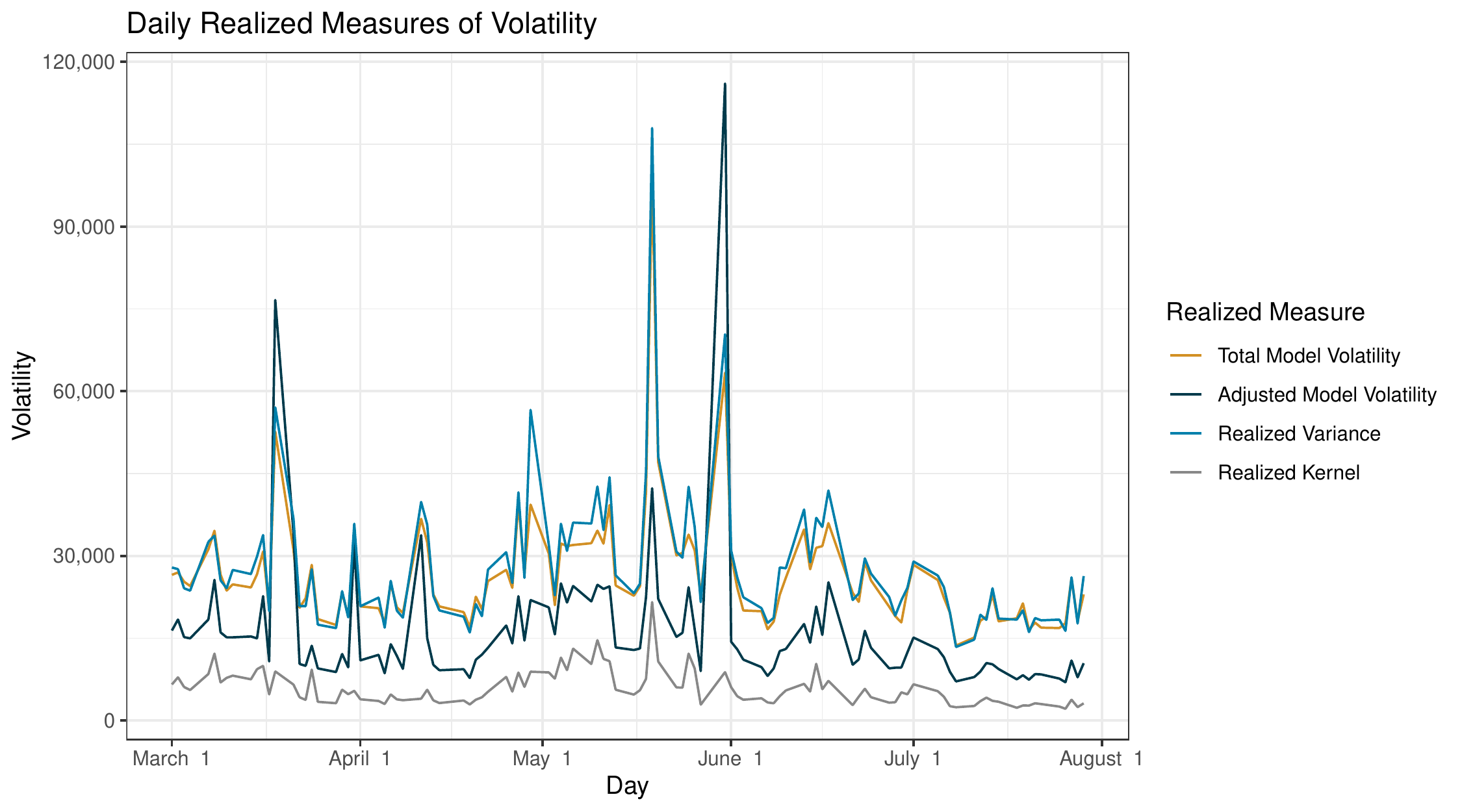}
\caption{The daily values of various volatility realized measures. The results are for the CSCO stock.}
\label{fig:realizedCSCO}
\end{figure}

\begin{figure}
\centering
\includegraphics[width=15cm]{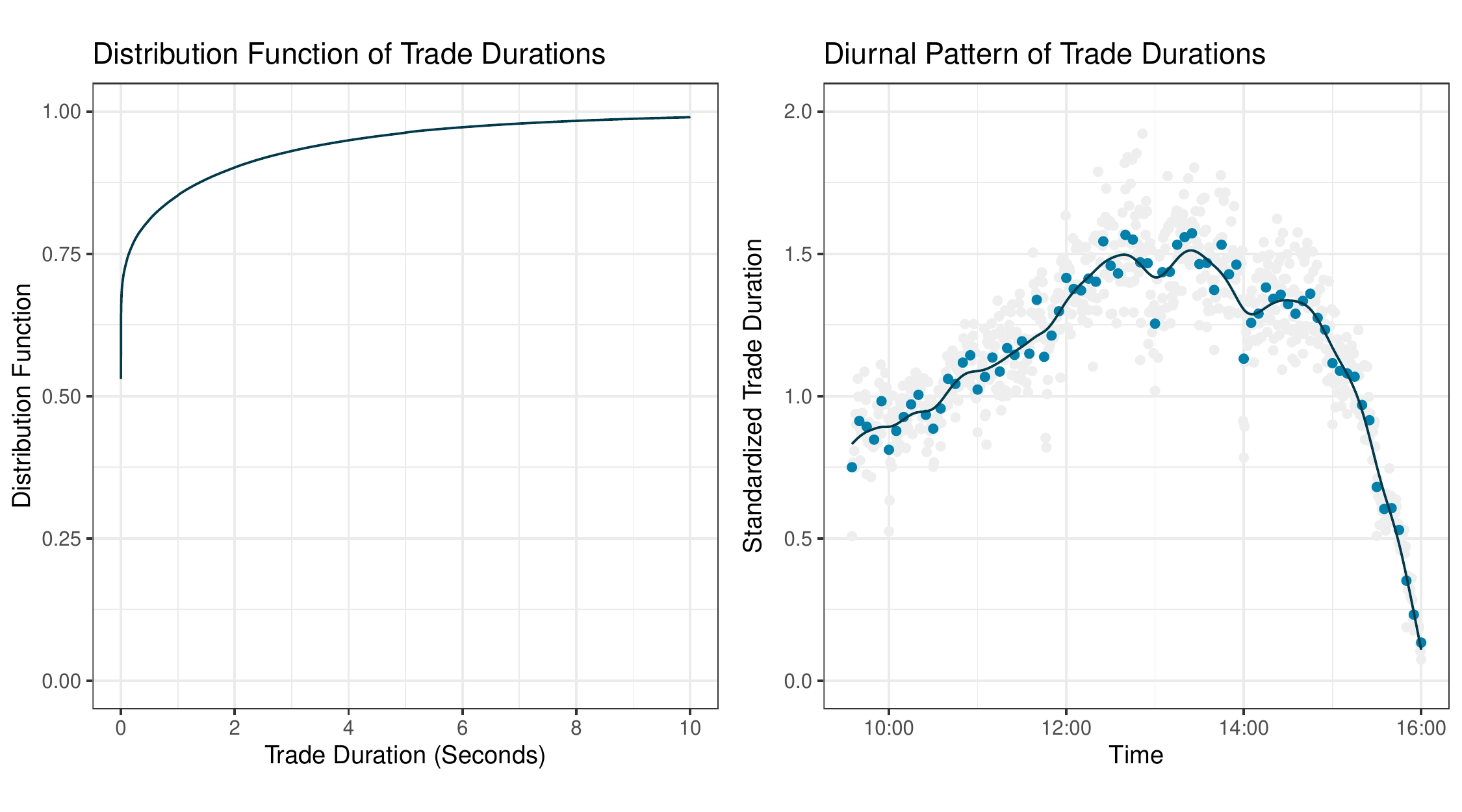}
\caption{Left: The empirical distribution function of trade durations. Right: The average trade durations in 5 minute (blue dots) and 30 second (grey dots) intraday intervals with a smoothed curve $\hat{f}_{\mathrm{dur}}(t_i)$. The results are for the EA stock.}
\label{fig:durationEA}
\end{figure}

\begin{figure}
\centering
\includegraphics[width=15cm]{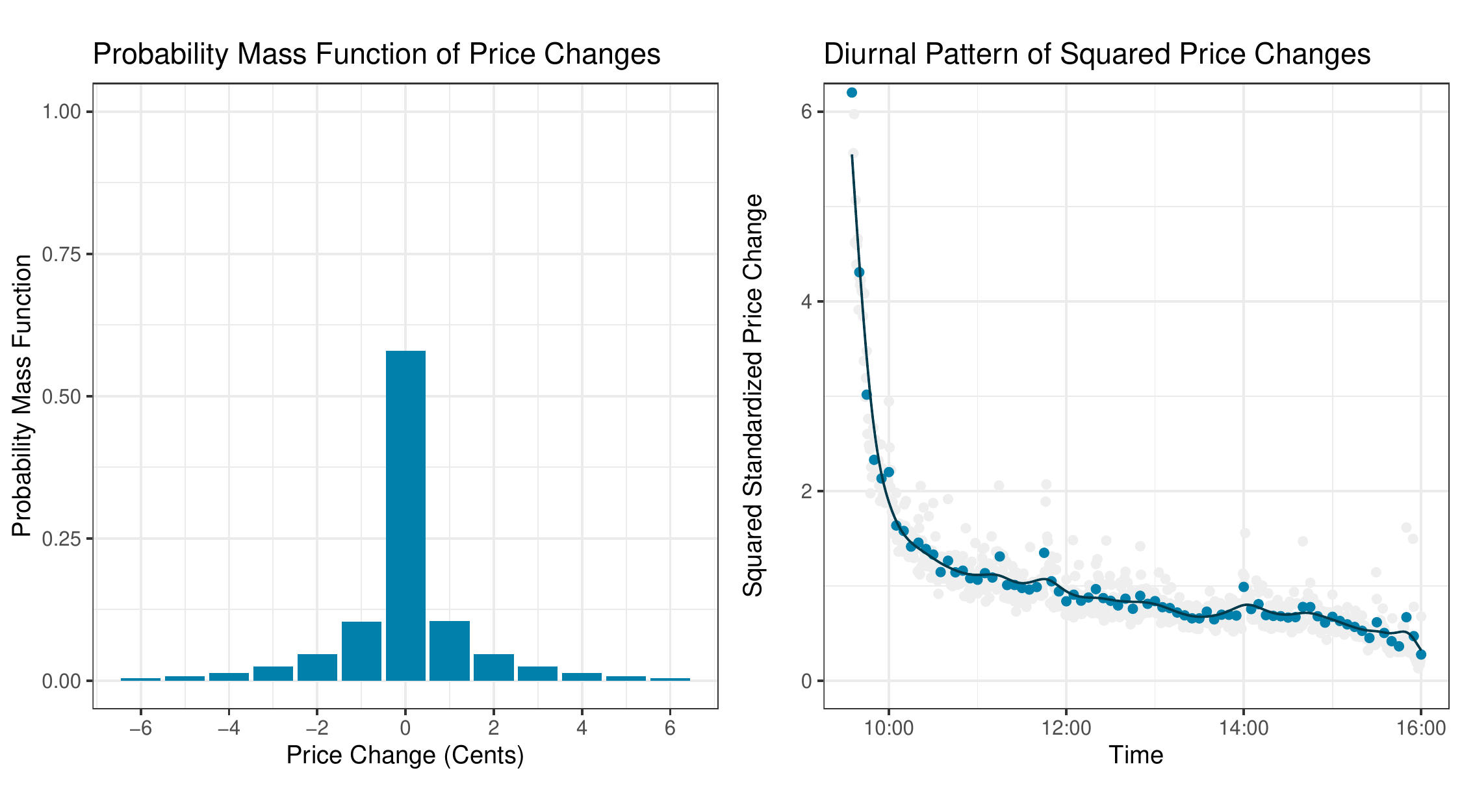}
\caption{Left: The empirical probability mass function of price changes. Right: The average squared price changes in 5 minute (blue dots) and 30 second (grey dots) intraday intervals with a smoothed curve $\hat{f}_{\mathrm{var}}(t_i)$. The results are for the EA stock.}
\label{fig:returnEA}
\end{figure}

\begin{figure}
\centering
\includegraphics[width=15cm]{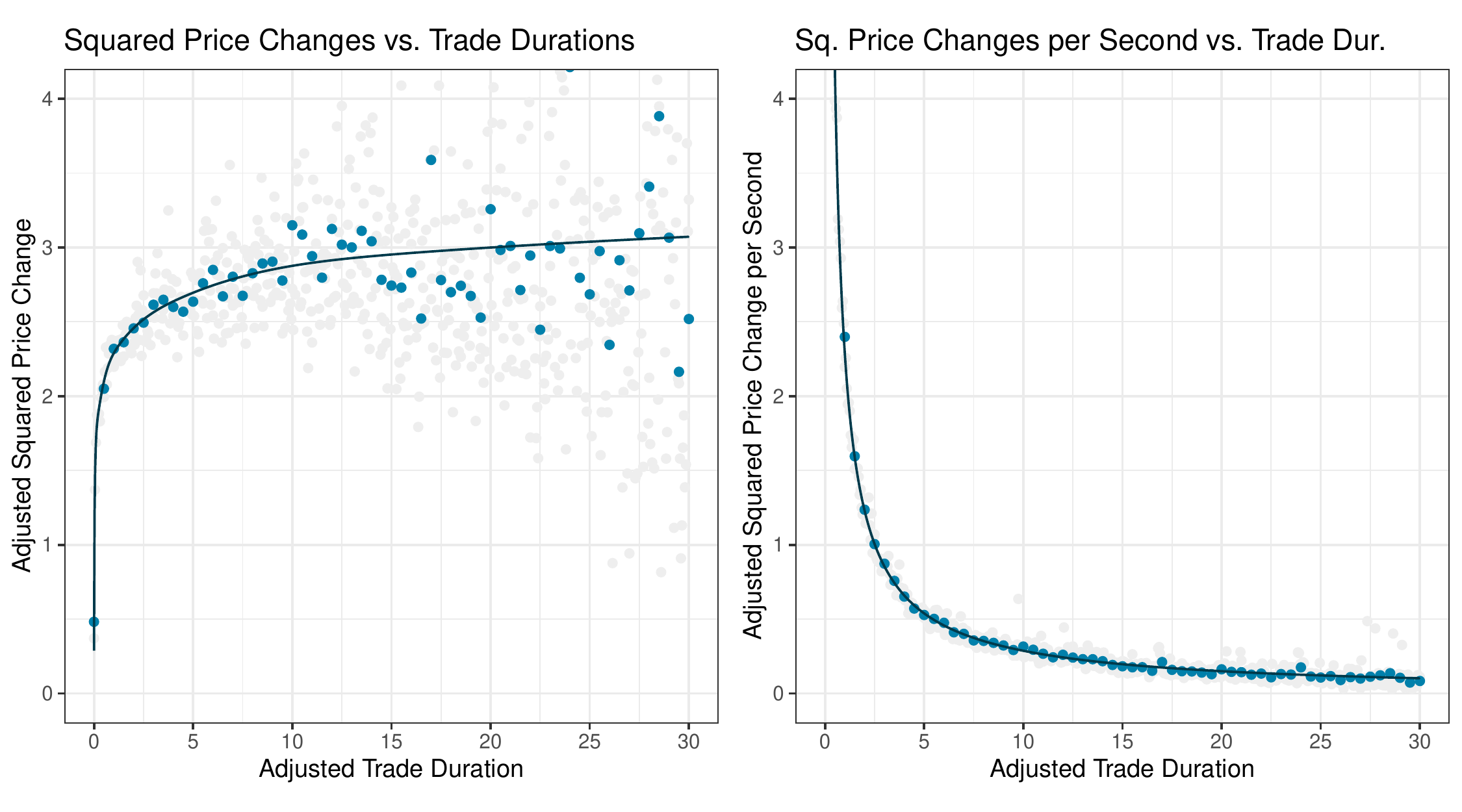}
\caption{Left: The average diurnally adjusted squared price changes in half second (blue dots) and 50 millisecond (grey dots) intervals of diurnally adjusted trade durations with a smoothed curve $\hat{f}_{\mathrm{rel}}(\tilde{d}_i)$. Right: The average diurnally adjusted squared price changes per second in half second (blue dots) and 50 millisecond (grey dots) intervals of diurnally adjusted trade durations with a smoothed curve $\hat{f}_{\mathrm{rel}}(\tilde{d}_i) / \tilde{d}_i$. The results are for the EA stock.}
\label{fig:relationEA}
\end{figure}

\begin{table}
\centering
\caption{The minimum, median, and maximum values of the estimated parameters of the models estimated on daily basis. The results are for the EA stock.}
\label{tab:coefEA}
\footnotesize
\begin{tabular}{llrrrrr}
\toprule
& & \multicolumn{5}{c}{Model} \\
\cmidrule(l{3pt}r{3pt}){3-7}
Coef. & Trans. & \multicolumn{1}{c}{Naive} & \multicolumn{1}{c}{No Infl.} & \multicolumn{1}{c}{Static Disp.} & \multicolumn{1}{c}{Static Mean} & \multicolumn{1}{c}{Proposed} \\
\midrule
          & Min &  & -0.369 & -0.636 &  & -0.571 \\ 
$\theta$  & Med &  & -0.247 & -0.426 &  & -0.365 \\ 
          & Max &  & -0.135 & -0.255 &  & -0.190 \\ \\
          & Min & 0.262 & -0.287 & 0.159 & 0.415 & 0.093 \\ 
$\omega$  & Med & 0.878 & 0.543 & 0.936 & 1.091 & 0.849 \\ 
          & Max & 2.135 & 1.476 & 2.393 & 2.376 & 2.644 \\ \\
          & Min &  & 0.838 &  & 0.617 & 0.900 \\ 
$\varphi$ & Med &  & 0.951 &  & 0.950 & 0.971 \\ 
          & Max &  & 0.998 &  & 1.000 & 0.999 \\ \\
          & Min &  & 0.032 &  & 0.010 & 0.033 \\ 
$\alpha$  & Med &  & 0.183 &  & 0.184 & 0.189 \\ 
          & Max &  & 0.346 &  & 0.491 & 0.347 \\ \\
          & Min &  &  & 0.165 & 0.137 & 0.153 \\ 
$\pi$     & Med &  &  & 0.307 & 0.264 & 0.274 \\ 
          & Max &  &  & 0.441 & 0.415 & 0.394 \\  
\bottomrule
\end{tabular}
\end{table}

\begin{table}
\centering
\caption{The R$^2$ statistics of the residuals and the squared residuals regressed on their lagged values with the average log-likelihood of an observation for the models estimated on daily basis. The results are for the EA stock.}
\label{tab:fitEA}
\footnotesize
\begin{tabular}{llrrrrr}
\toprule
& & \multicolumn{5}{c}{Model} \\
\cmidrule(l{3pt}r{3pt}){3-7}
Statistic & Lag & \multicolumn{1}{c}{Naive} & \multicolumn{1}{c}{No Infl.} & \multicolumn{1}{c}{Static Disp.} & \multicolumn{1}{c}{Static Mean} & \multicolumn{1}{c}{Proposed} \\
\midrule
               &   1 & 0.101 & 0.006 & 0.003 & 0.074 & 0.004 \\ 
AR R$^2$       &  10 & 0.134 & 0.013 & 0.005 & 0.093 & 0.007 \\ 
               & 100 & 0.141 & 0.017 & 0.009 & 0.096 & 0.010 \\ \\
               &   1 & 0.094 & 0.001 & 0.002 & 0.013 & 0.001 \\ 
ARCH R$^2$     &  10 & 0.136 & 0.003 & 0.028 & 0.015 & 0.003 \\ 
               & 100 & 0.170 & 0.011 & 0.051 & 0.022 & 0.009 \\ \\
Log-Likelihood &     & -1.590 & -1.473 & -1.457 & -1.461 & -1.424 \\ 
\bottomrule
\end{tabular}
\end{table}

\begin{table}
\centering
\caption{The R$^2$ statistics of the residuals and the squared residuals regressed on their lagged values with the average log-likelihood of an observation for the alternative models estimated on daily basis. The results are for the EA stock.}
\label{tab:altEA}
\footnotesize
\begin{tabular}{llrrrrr}
\toprule
& & \multicolumn{5}{c}{Model} \\
\cmidrule(l{3pt}r{3pt}){3-7}
Statistic & Lag & \multicolumn{1}{c}{Var. Param.} & \multicolumn{1}{c}{GAS Mean} & \multicolumn{1}{c}{GAS Infl.} & \multicolumn{1}{c}{Normal} & \multicolumn{1}{c}{Student's-t} \\
\midrule
               &   1 & 0.045 & 0.029 & 0.003 & 0.008 & NA \\ 
AR R$^2$       &  10 & 0.060 & 0.036 & 0.007 & 0.016 & NA \\ 
               & 100 & 0.063 & 0.040 & 0.010 & 0.019 & NA \\ \\
               &   1 & 0.009 & 0.004 & 0.000 & 0.001 & NA \\ 
ARCH R$^2$     &  10 & 0.012 & 0.007 & 0.002 & 0.001 & NA \\ 
               & 100 & 0.019 & 0.014 & 0.008 & 0.008 & NA \\ \\
Log-Lik.       &     & -1.446 & -1.445 & -1.419 & -1.568 & -3.347 \\ 
\bottomrule
\end{tabular}
\end{table}

\begin{table}
\centering
\caption{The average log-likelihood, mean absolute error (MAE), and root mean squared error (RMSE) in the test sample for the naive and proposed models estimated on daily basis. The results are for the EA stock.}
\label{tab:fcstEA}
\footnotesize
\begin{tabular}{llrrrr}
\toprule
& & \multicolumn{2}{c}{Full Train Sample} & \multicolumn{2}{c}{Agg. Train Sample} \\
\cmidrule(l{3pt}r{3pt}){3-4} \cmidrule(l{3pt}r{3pt}){5-6}
Test Sample & Statistic & \multicolumn{1}{c}{Naive} & \multicolumn{1}{c}{Proposed} & \multicolumn{1}{c}{Naive} & \multicolumn{1}{c}{Proposed} \\
\midrule
           & Log-Lik. & -1.599 & -1.425 & -1.725 & -1.459 \\ 
Full       & MAE      &  0.933 &  0.986 &  0.933 &  0.990 \\ 
           & RMSE     &  1.969 &  1.846 &  1.969 &  1.843 \\ \\
           & Log-Lik. & -2.352 & -2.041 & -2.170 & -2.008 \\ 
Aggregated & MAE      &  1.593 &  1.622 &  1.593 &  1.625 \\ 
           & RMSE     &  2.686 &  2.522 &  2.686 &  2.517 \\ 
\bottomrule
\end{tabular}
\end{table}

\begin{figure}
\centering
\includegraphics[width=15cm]{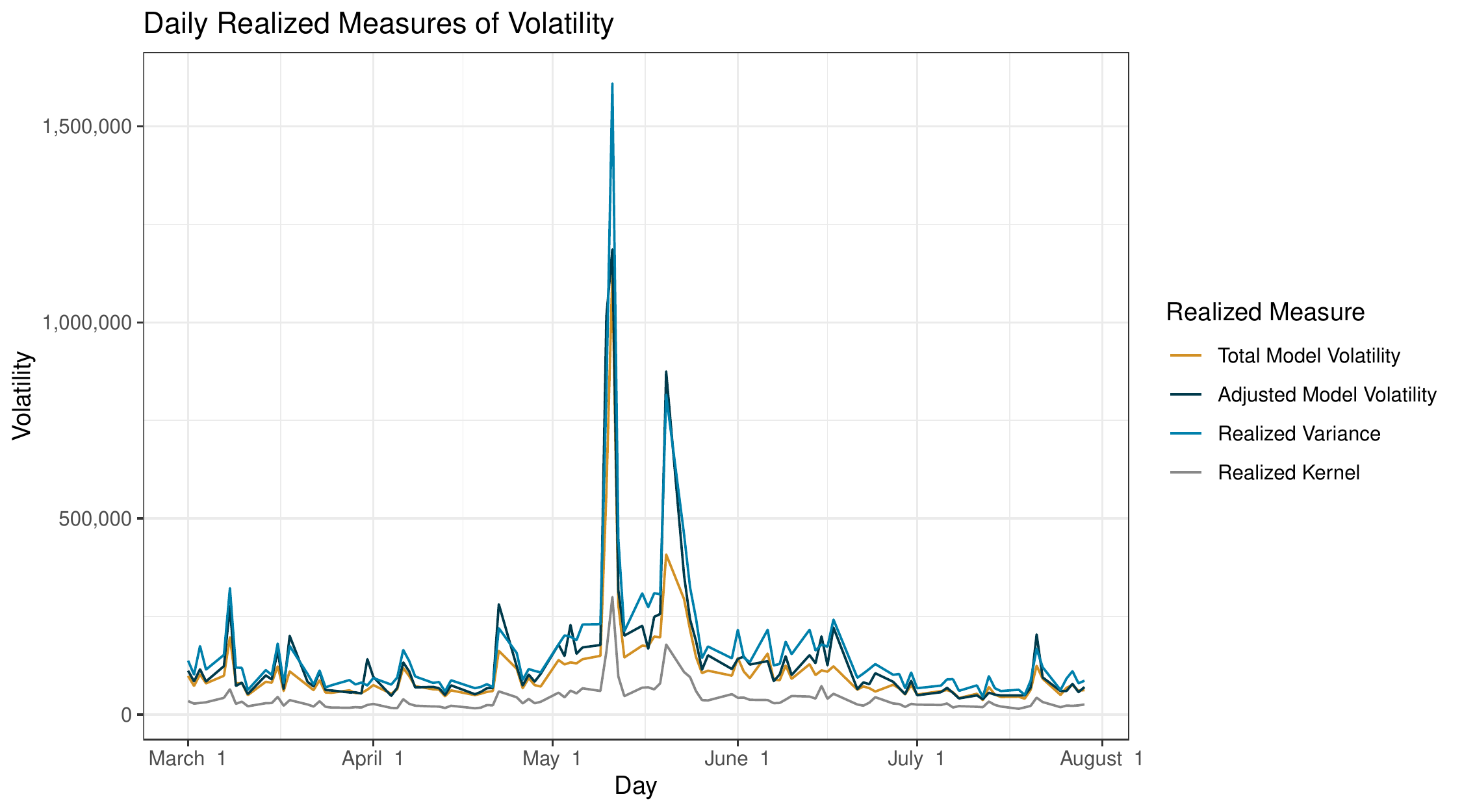}
\caption{The daily values of various volatility realized measures. The results are for the EA stock.}
\label{fig:realizedEA}
\end{figure}

\begin{figure}
\centering
\includegraphics[width=15cm]{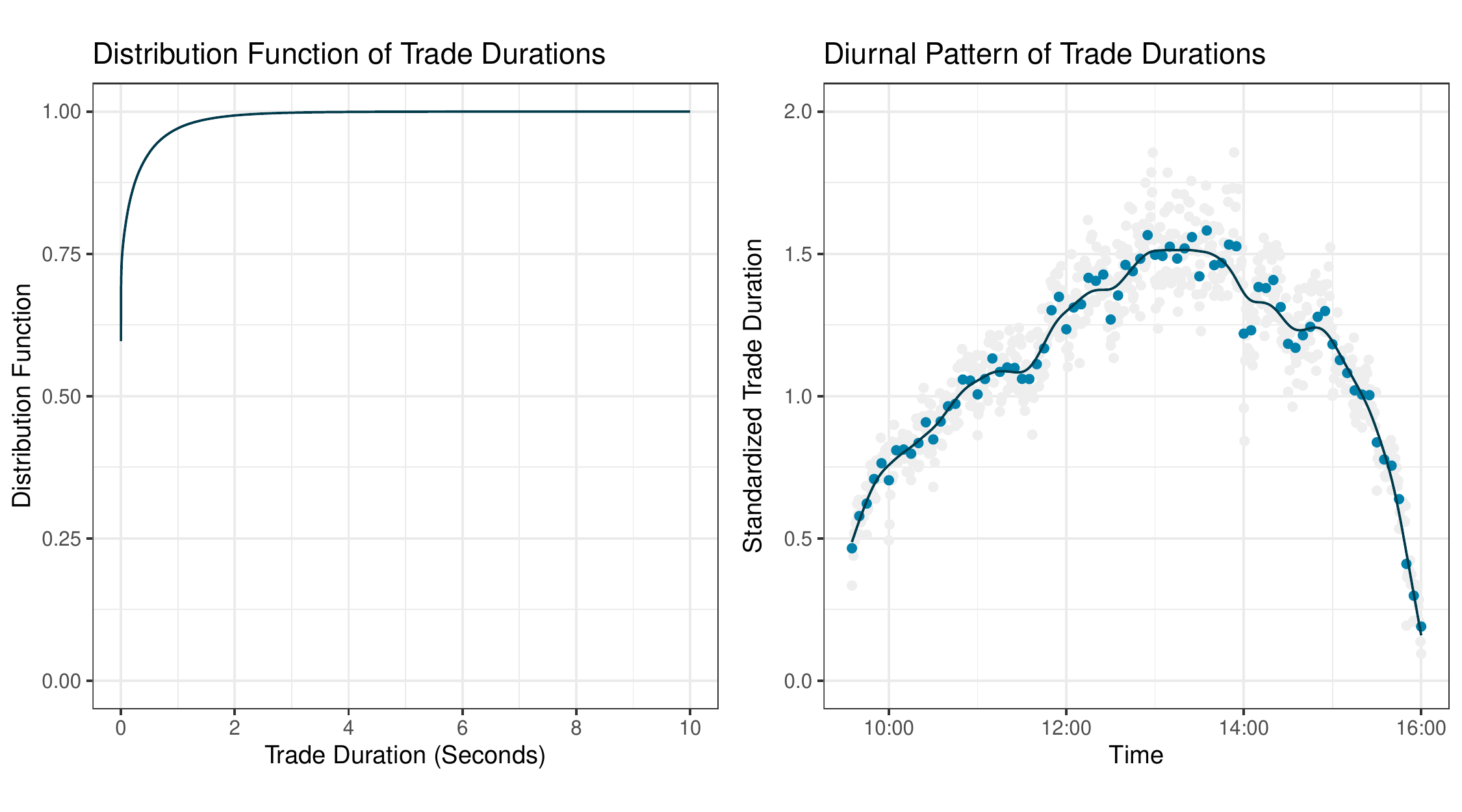}
\caption{Left: The empirical distribution function of trade durations. Right: The average trade durations in 5 minute (blue dots) and 30 second (grey dots) intraday intervals with a smoothed curve $\hat{f}_{\mathrm{dur}}(t_i)$. The results are for the INTC stock.}
\label{fig:durationINTC}
\end{figure}

\begin{figure}
\centering
\includegraphics[width=15cm]{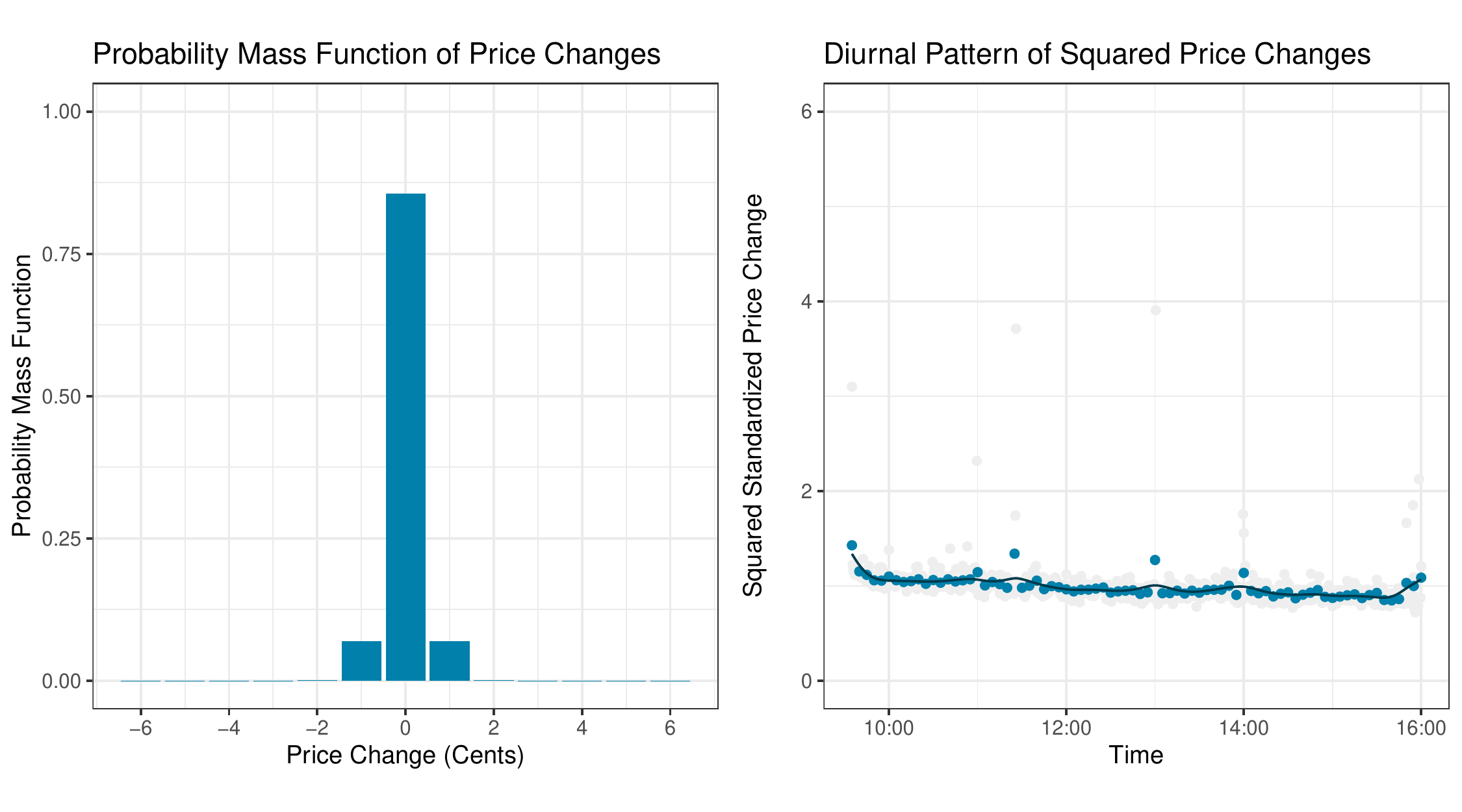}
\caption{Left: The empirical probability mass function of price changes. Right: The average squared price changes in 5 minute (blue dots) and 30 second (grey dots) intraday intervals with a smoothed curve $\hat{f}_{\mathrm{var}}(t_i)$. The results are for the INTC stock.}
\label{fig:returnINTC}
\end{figure}

\begin{figure}
\centering
\includegraphics[width=15cm]{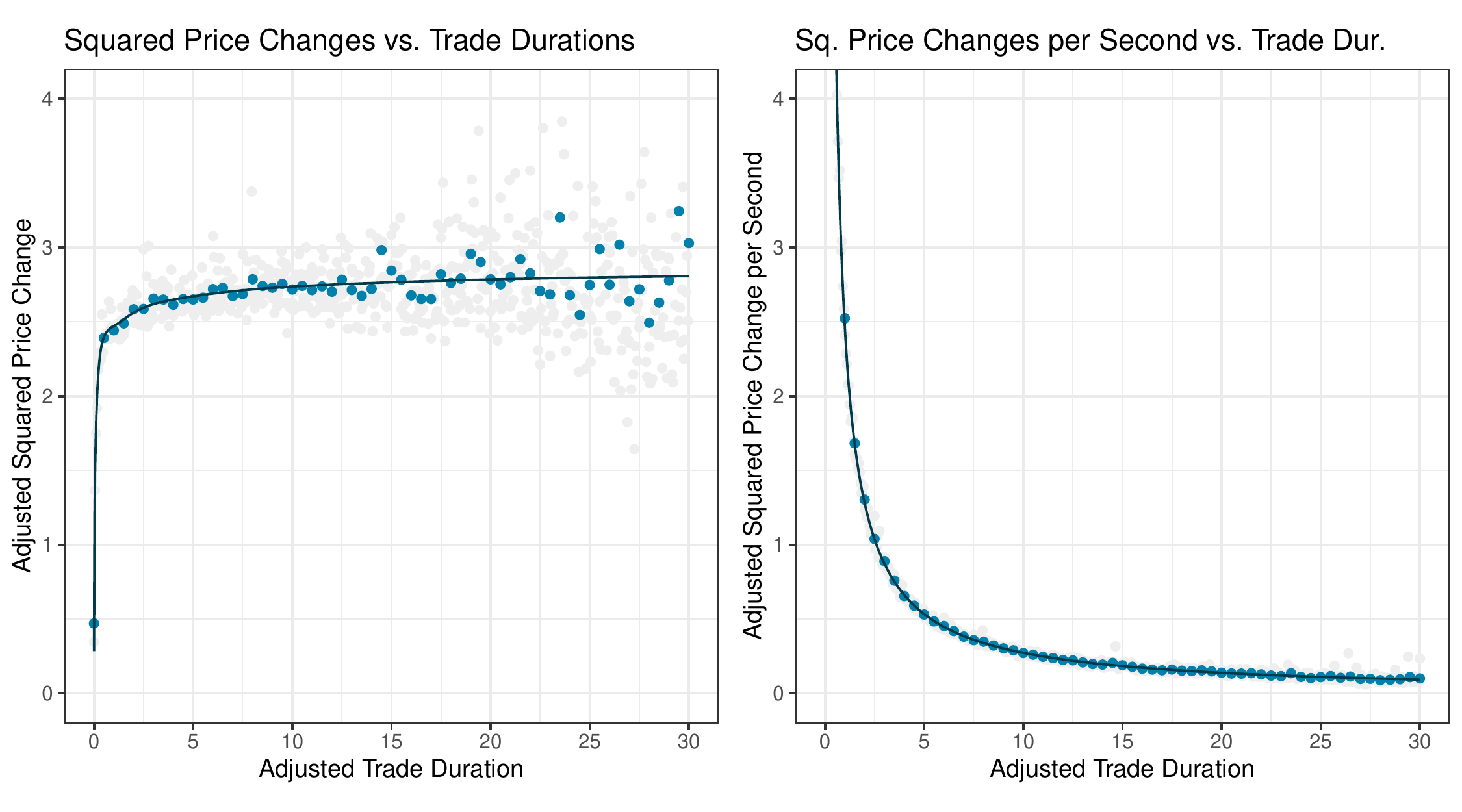}
\caption{Left: The average diurnally adjusted squared price changes in half second (blue dots) and 50 millisecond (grey dots) intervals of diurnally adjusted trade durations with a smoothed curve $\hat{f}_{\mathrm{rel}}(\tilde{d}_i)$. Right: The average diurnally adjusted squared price changes per second in half second (blue dots) and 50 millisecond (grey dots) intervals of diurnally adjusted trade durations with a smoothed curve $\hat{f}_{\mathrm{rel}}(\tilde{d}_i) / \tilde{d}_i$. The results are for the INTC stock.}
\label{fig:relationINTC}
\end{figure}

\begin{table}
\centering
\caption{The minimum, median, and maximum values of the estimated parameters of the models estimated on daily basis. The results are for the INTC stock.}
\label{tab:coefINTC}
\footnotesize
\begin{tabular}{llrrrrr}
\toprule
& & \multicolumn{5}{c}{Model} \\
\cmidrule(l{3pt}r{3pt}){3-7}
Coef. & Trans. & \multicolumn{1}{c}{Naive} & \multicolumn{1}{c}{No Infl.} & \multicolumn{1}{c}{Static Disp.} & \multicolumn{1}{c}{Static Mean} & \multicolumn{1}{c}{Proposed} \\
\midrule
          & Min &  & -0.694 & -0.695 &  & -0.694 \\ 
$\theta$  & Med &  & -0.519 & -0.523 &  & -0.519 \\ 
          & Max &  & -0.323 & -0.335 &  & -0.323 \\ \\
          & Min & -1.930 & -3.056 & -3.055 & -2.105 & -3.056 \\ 
$\omega$  & Med & -1.760 & -2.552 & -2.565 & -1.874 & -2.552 \\ 
          & Max & -1.570 & -0.755 & -2.186 & -1.619 & -0.815 \\ \\
          & Min &  & 0.964 &  & 0.740 & 0.963 \\ 
$\varphi$ & Med &  & 0.998 &  & 0.871 & 0.998 \\ 
          & Max &  & 1.000 &  & 0.993 & 1.000 \\ \\
          & Min &  & 0.006 &  & 0.104 & 0.006 \\ 
$\alpha$  & Med &  & 0.061 &  & 0.842 & 0.061 \\ 
          & Max &  & 0.197 &  & 1.018 & 0.197 \\ \\
          & Min &  &  & 0.000 & 0.000 & 0.000 \\ 
$\pi$     & Med &  &  & 0.000 & 0.000 & 0.000 \\ 
          & Max &  &  & 0.021 & 0.097 & 0.016 \\ 
\bottomrule
\end{tabular}
\end{table}

\begin{table}
\centering
\caption{The R$^2$ statistics of the residuals and the squared residuals regressed on their lagged values with the average log-likelihood of an observation for the models estimated on daily basis. The results are for the INTC stock.}
\label{tab:fitINTC}
\footnotesize
\begin{tabular}{llrrrrr}
\toprule
& & \multicolumn{5}{c}{Model} \\
\cmidrule(l{3pt}r{3pt}){3-7}
Statistic & Lag & \multicolumn{1}{c}{Naive} & \multicolumn{1}{c}{No Infl.} & \multicolumn{1}{c}{Static Disp.} & \multicolumn{1}{c}{Static Mean} & \multicolumn{1}{c}{Proposed} \\
\midrule
               &   1 & 0.132 & 0.000 & 0.000 & 0.079 & 0.000 \\ 
AR R$^2$       &  10 & 0.186 & 0.003 & 0.003 & 0.109 & 0.003 \\ 
               & 100 & 0.188 & 0.004 & 0.004 & 0.109 & 0.004 \\ \\
               &   1 & 0.096 & 0.000 & 0.000 & 0.008 & 0.000 \\ 
ARCH R$^2$     &  10 & 0.143 & 0.002 & 0.003 & 0.010 & 0.002 \\ 
               & 100 & 0.175 & 0.008 & 0.019 & 0.022 & 0.008 \\ \\               
Log-Likelihood &     & -0.464 & -0.399 & -0.400 & -0.442 & -0.399 \\ 
\bottomrule
\end{tabular}
\end{table}

\begin{table}
\centering
\caption{The R$^2$ statistics of the residuals and the squared residuals regressed on their lagged values with the average log-likelihood of an observation for the alternative models estimated on daily basis. The results are for the INTC stock.}
\label{tab:altINTC}
\footnotesize
\begin{tabular}{llrrrrr}
\toprule
& & \multicolumn{5}{c}{Model} \\
\cmidrule(l{3pt}r{3pt}){3-7}
Statistic & Lag & \multicolumn{1}{c}{Var. Param.} & \multicolumn{1}{c}{GAS Mean} & \multicolumn{1}{c}{GAS Infl.} & \multicolumn{1}{c}{Normal} & \multicolumn{1}{c}{Student's-t} \\
\midrule
               &   1 & 0.061 & 0.022 & 0.000 & 0.004 & NA \\ 
AR R$^2$       &  10 & 0.090 & 0.031 & 0.003 & 0.013 & NA \\ 
               & 100 & 0.091 & 0.035 & 0.004 & 0.014 & NA \\ \\
               &   1 & 0.005 & 0.001 & 0.000 & 0.000 & NA \\ 
ARCH R$^2$     &  10 & 0.012 & 0.013 & 0.002 & 0.000 & NA \\ 
               & 100 & 0.029 & 0.022 & 0.009 & 0.002 & NA \\ \\        
Log-Lik.       &     &  -0.435 & -0.426 & -0.399 & -0.423 & -1.306 \\ 
\bottomrule
\end{tabular}
\end{table}

\begin{table}
\centering
\caption{The average log-likelihood, mean absolute error (MAE), and root mean squared error (RMSE) in the test sample for the naive and proposed models estimated on daily basis. The results are for the INTC stock.}
\label{tab:fcstINTC}
\footnotesize
\begin{tabular}{llrrrr}
\toprule
& & \multicolumn{2}{c}{Full Train Sample} & \multicolumn{2}{c}{Agg. Train Sample} \\
\cmidrule(l{3pt}r{3pt}){3-4} \cmidrule(l{3pt}r{3pt}){5-6}
Test Sample & Statistic & \multicolumn{1}{c}{Naive} & \multicolumn{1}{c}{Proposed} & \multicolumn{1}{c}{Naive} & \multicolumn{1}{c}{Proposed} \\
\midrule
           & Log-Lik. & -0.464 & -0.401 & -0.507 & -0.407 \\ 
Full       & MAE      &  0.149 &  0.178 &  0.149 &  0.177 \\ 
           & RMSE     &  0.415 &  0.373 &  0.415 &  0.373 \\ \\
           & Log-Lik. & -0.910 & -0.766 & -0.847 & -0.763 \\ 
Aggregated & MAE      &  0.309 &  0.355 &  0.309 &  0.354 \\ 
           & RMSE     &  0.597 &  0.538 &  0.597 &  0.538 \\ 
\bottomrule
\end{tabular}
\end{table}

\begin{figure}
\centering
\includegraphics[width=15cm]{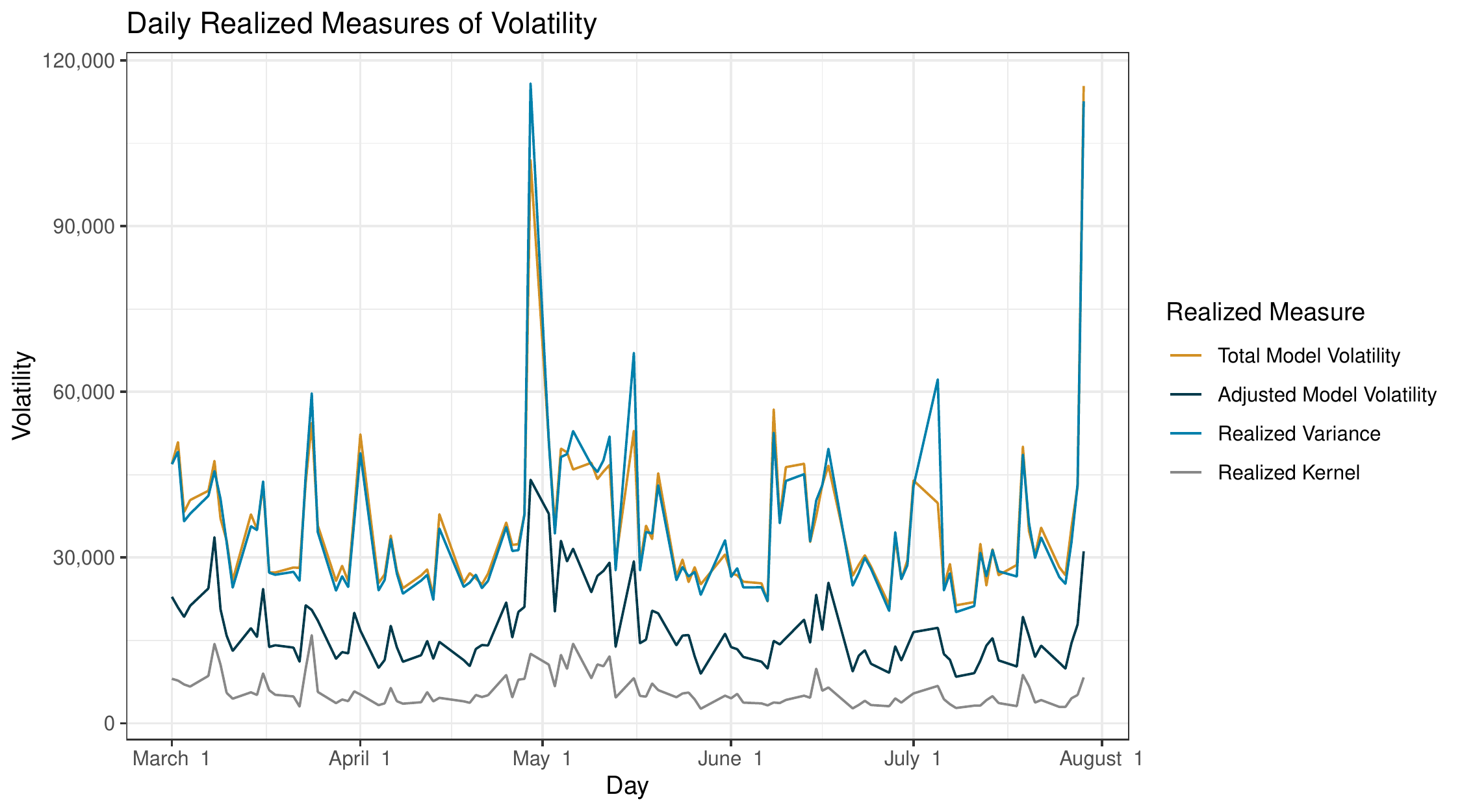}
\caption{The daily values of various volatility realized measures. The results are for the INTC stock.}
\label{fig:realizedINTC}
\end{figure}

\begin{figure}
\centering
\includegraphics[width=15cm]{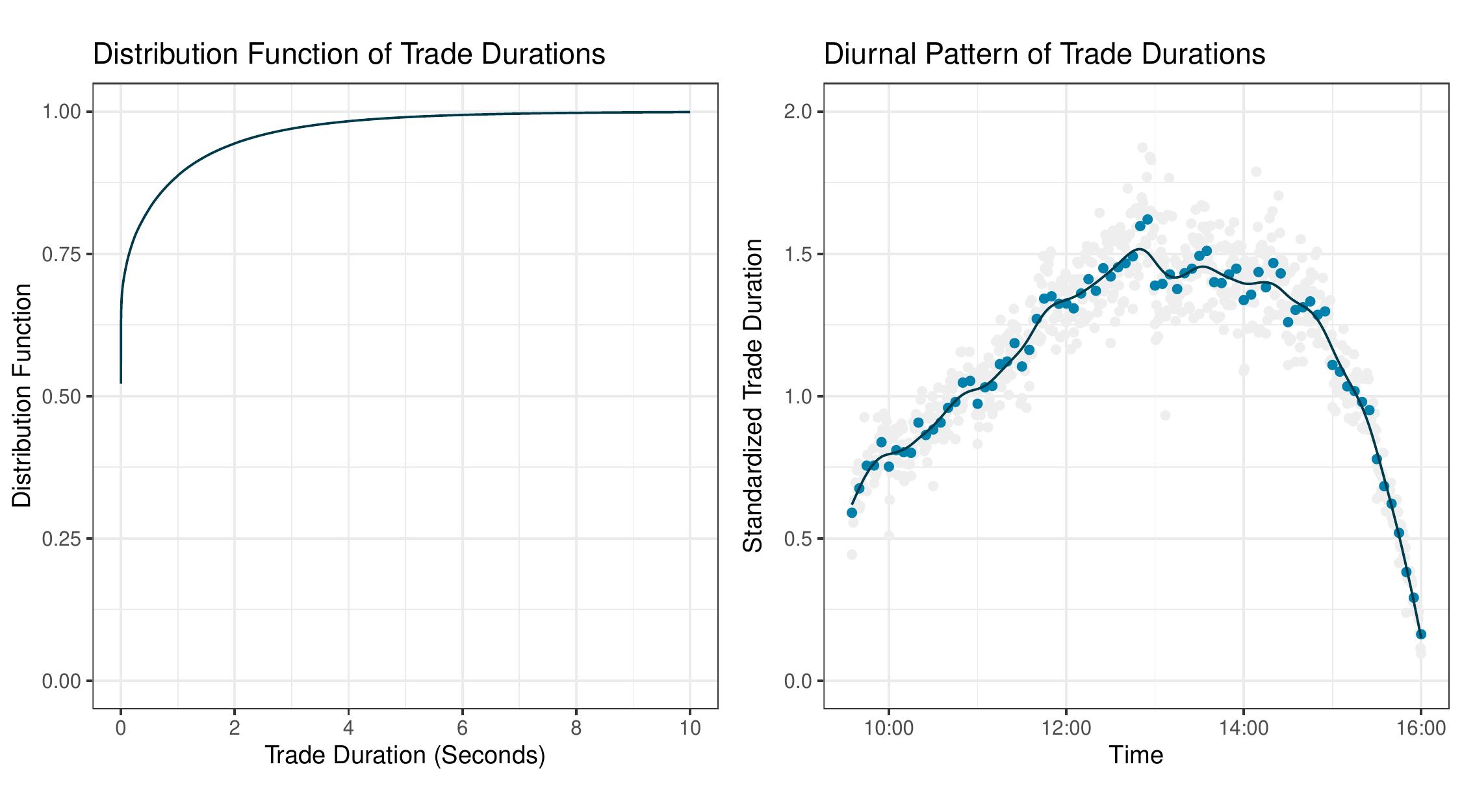}
\caption{Left: The empirical distribution function of trade durations. Right: The average trade durations in 5 minute (blue dots) and 30 second (grey dots) intraday intervals with a smoothed curve $\hat{f}_{\mathrm{dur}}(t_i)$. The results are for the MA stock.}
\label{fig:durationMA}
\end{figure}

\begin{figure}
\centering
\includegraphics[width=15cm]{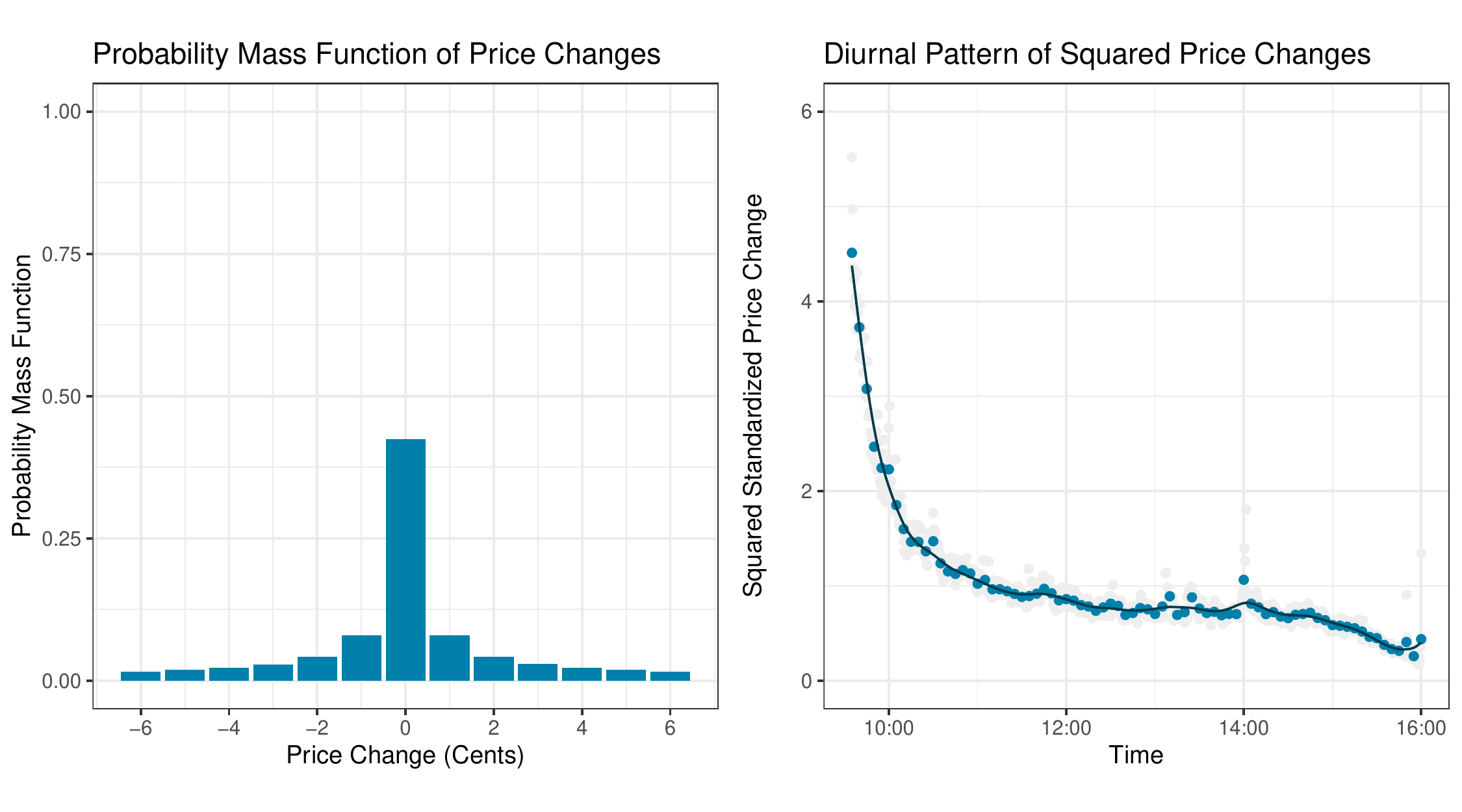}
\caption{Left: The empirical probability mass function of price changes. Right: The average squared price changes in 5 minute (blue dots) and 30 second (grey dots) intraday intervals with a smoothed curve $\hat{f}_{\mathrm{var}}(t_i)$. The results are for the MA stock.}
\label{fig:returnMA}
\end{figure}

\begin{figure}
\centering
\includegraphics[width=15cm]{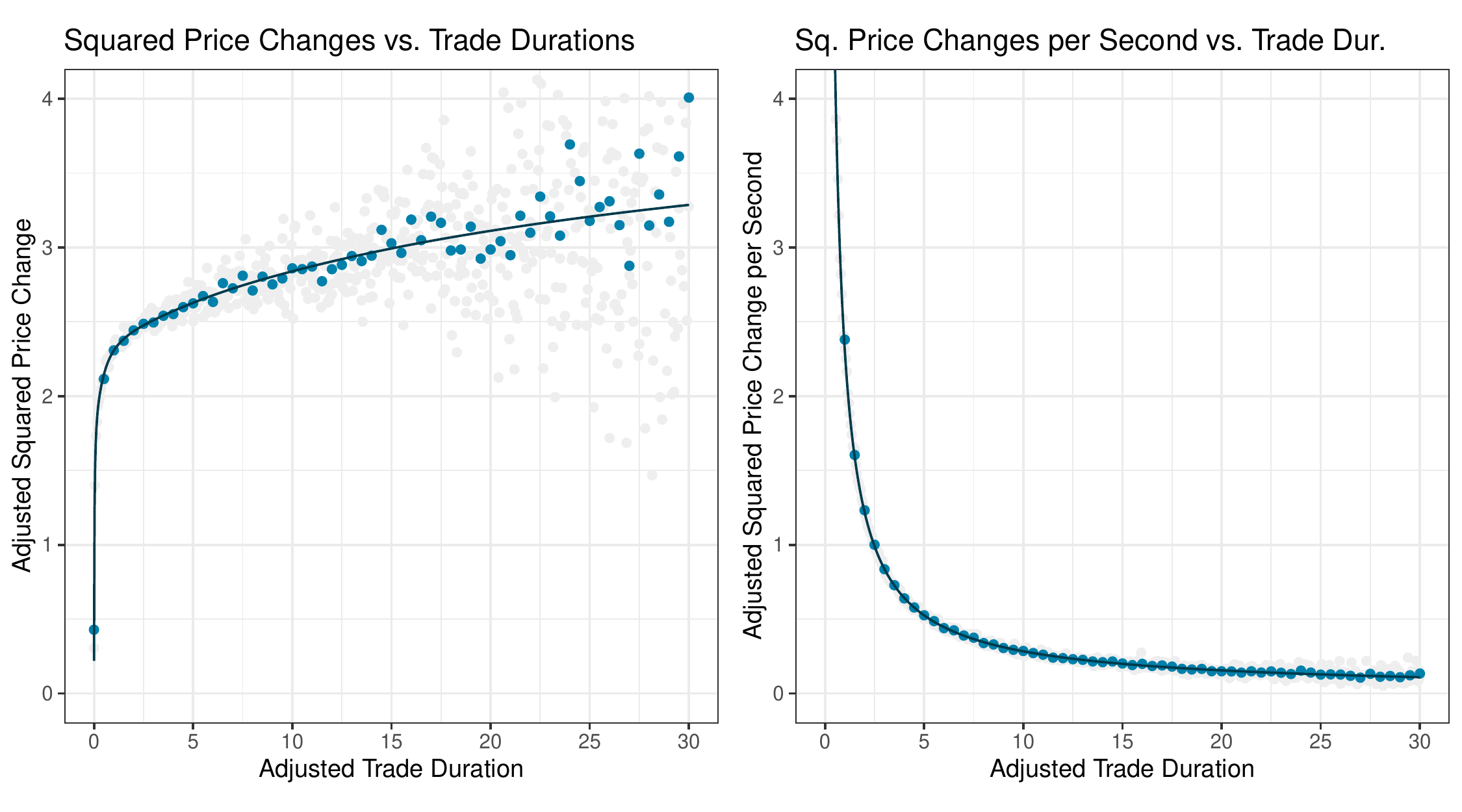}
\caption{Left: The average diurnally adjusted squared price changes in half second (blue dots) and 50 millisecond (grey dots) intervals of diurnally adjusted trade durations with a smoothed curve $\hat{f}_{\mathrm{rel}}(\tilde{d}_i)$. Right: The average diurnally adjusted squared price changes per second in half second (blue dots) and 50 millisecond (grey dots) intervals of diurnally adjusted trade durations with a smoothed curve $\hat{f}_{\mathrm{rel}}(\tilde{d}_i) / \tilde{d}_i$. The results are for the MA stock.}
\label{fig:relationMA}
\end{figure}

\begin{table}
\centering
\caption{The minimum, median, and maximum values of the estimated parameters of the models estimated on daily basis. The results are for the MA stock.}
\label{tab:coefMA}
\footnotesize
\begin{tabular}{llrrrrr}
\toprule
& & \multicolumn{5}{c}{Model} \\
\cmidrule(l{3pt}r{3pt}){3-7}
Coef. & Trans. & \multicolumn{1}{c}{Naive} & \multicolumn{1}{c}{No Infl.} & \multicolumn{1}{c}{Static Disp.} & \multicolumn{1}{c}{Static Mean} & \multicolumn{1}{c}{Proposed} \\
\midrule
          & Min &  & -0.398 & -0.555 &  & -0.545 \\ 
$\theta$  & Med &  & -0.315 & -0.468 &  & -0.449 \\ 
          & Max &  & -0.231 & -0.397 &  & -0.358 \\ \\
          & Min & 2.495 & 2.010 & 2.633 & 1.934 & 1.935 \\ 
$\omega$  & Med & 3.218 & 2.690 & 3.372 & 3.471 & 3.177 \\ 
          & Max & 3.974 & 3.800 & 3.963 & 3.947 & 3.930 \\ \\
          & Min &  & 0.829 &  & 0.442 & 0.886 \\ 
$\varphi$ & Med &  & 0.978 &  & 0.986 & 0.991 \\ 
          & Max &  & 1.000 &  & 1.000 & 1.000 \\ \\
          & Min &  & 0.001 &  & 0.000 & 0.000 \\ 
$\alpha$  & Med &  & 0.053 &  & 0.023 & 0.034 \\ 
          & Max &  & 0.168 &  & 0.149 & 0.154 \\ \\
          & Min &  &  & 0.249 & 0.230 & 0.233 \\ 
$\pi$     & Med &  &  & 0.333 & 0.324 & 0.320 \\ 
          & Max &  &  & 0.391 & 0.392 & 0.387 \\ 
\bottomrule
\end{tabular}
\end{table}

\begin{table}
\centering
\caption{The R$^2$ statistics of the residuals and the squared residuals regressed on their lagged values with the average log-likelihood of an observation for the models estimated on daily basis. The results are for the MA stock.}
\label{tab:fitMA}
\footnotesize
\begin{tabular}{llrrrrr}
\toprule
& & \multicolumn{5}{c}{Model} \\
\cmidrule(l{3pt}r{3pt}){3-7}
Statistic & Lag & \multicolumn{1}{c}{Naive} & \multicolumn{1}{c}{No Infl.} & \multicolumn{1}{c}{Static Disp.} & \multicolumn{1}{c}{Static Mean} & \multicolumn{1}{c}{Proposed} \\
\midrule
               &   1 & 0.125 & 0.005 & 0.008 & 0.118 & 0.008 \\ 
AR R$^2$       &  10 & 0.167 & 0.015 & 0.011 & 0.157 & 0.011 \\ 
               & 100 & 0.170 & 0.017 & 0.013 & 0.159 & 0.013 \\ \\
               &   1 & 0.090 & 0.010 & 0.005 & 0.057 & 0.004 \\ 
ARCH R$^2$     &  10 & 0.122 & 0.020 & 0.037 & 0.069 & 0.019 \\ 
               & 100 & 0.149 & 0.024 & 0.059 & 0.078 & 0.025 \\ \\
Log-Likelihood &     & -2.736 & -2.588 & -2.409 & -2.448 & -2.378 \\
\bottomrule
\end{tabular}
\end{table}

\begin{table}
\centering
\caption{The R$^2$ statistics of the residuals and the squared residuals regressed on their lagged values with the average log-likelihood of an observation for the alternative models estimated on daily basis. The results are for the MA stock.}
\label{tab:altMA}
\footnotesize
\begin{tabular}{llrrrrr}
\toprule
& & \multicolumn{5}{c}{Model} \\
\cmidrule(l{3pt}r{3pt}){3-7}
Statistic & Lag & \multicolumn{1}{c}{Var. Param.} & \multicolumn{1}{c}{GAS Mean} & \multicolumn{1}{c}{GAS Infl.} & \multicolumn{1}{c}{Normal} & \multicolumn{1}{c}{Student's-t} \\
\midrule
               &   1 & 0.061 & 0.051 & 0.004 & 0.013 & NA \\ 
AR R$^2$       &  10 & 0.087 & 0.061 & 0.009 & 0.024 & NA \\ 
               & 100 & 0.088 & 0.063 & 0.011 & 0.026 & NA \\ \\
               &   1 & 0.030 & 0.017 & 0.002 & 0.001 & NA \\ 
ARCH R$^2$     &  10 & 0.044 & 0.034 & 0.013 & 0.002 & NA \\ 
               & 100 & 0.053 & 0.043 & 0.019 & 0.006 & NA \\ \\
Log-Lik.       &     & -2.421 & -2.419 & -2.341 & -2.618 & -4.855 \\ 
\bottomrule
\end{tabular}
\end{table}

\begin{table}
\centering
\caption{The average log-likelihood, mean absolute error (MAE), and root mean squared error (RMSE) in the test sample for the naive and proposed models estimated on daily basis. The results are for the MA stock.}
\label{tab:fcstMA}
\footnotesize
\begin{tabular}{llrrrr}
\toprule
& & \multicolumn{2}{c}{Full Train Sample} & \multicolumn{2}{c}{Agg. Train Sample} \\
\cmidrule(l{3pt}r{3pt}){3-4} \cmidrule(l{3pt}r{3pt}){5-6}
Test Sample & Statistic & \multicolumn{1}{c}{Naive} & \multicolumn{1}{c}{Proposed} & \multicolumn{1}{c}{Naive} & \multicolumn{1}{c}{Proposed} \\
\midrule
           & Log-Lik. & -2.760 & -2.375 & -2.839 & -2.435 \\ 
Full       & MAE      &  3.006 &  3.037 &  3.006 &  3.059 \\ 
           & RMSE     &  5.743 &  5.243 &  5.743 &  5.196 \\ \\
           & Log-Lik. & -3.553 & -3.158 & -3.313 & -3.068 \\ 
Aggregated & MAE      &  5.162 &  5.022 &  5.162 &  5.028 \\ 
           & RMSE     &  8.030 &  7.301 &  8.030 &  7.227 \\            
\bottomrule
\end{tabular}
\end{table}

\begin{figure}
\centering
\includegraphics[width=15cm]{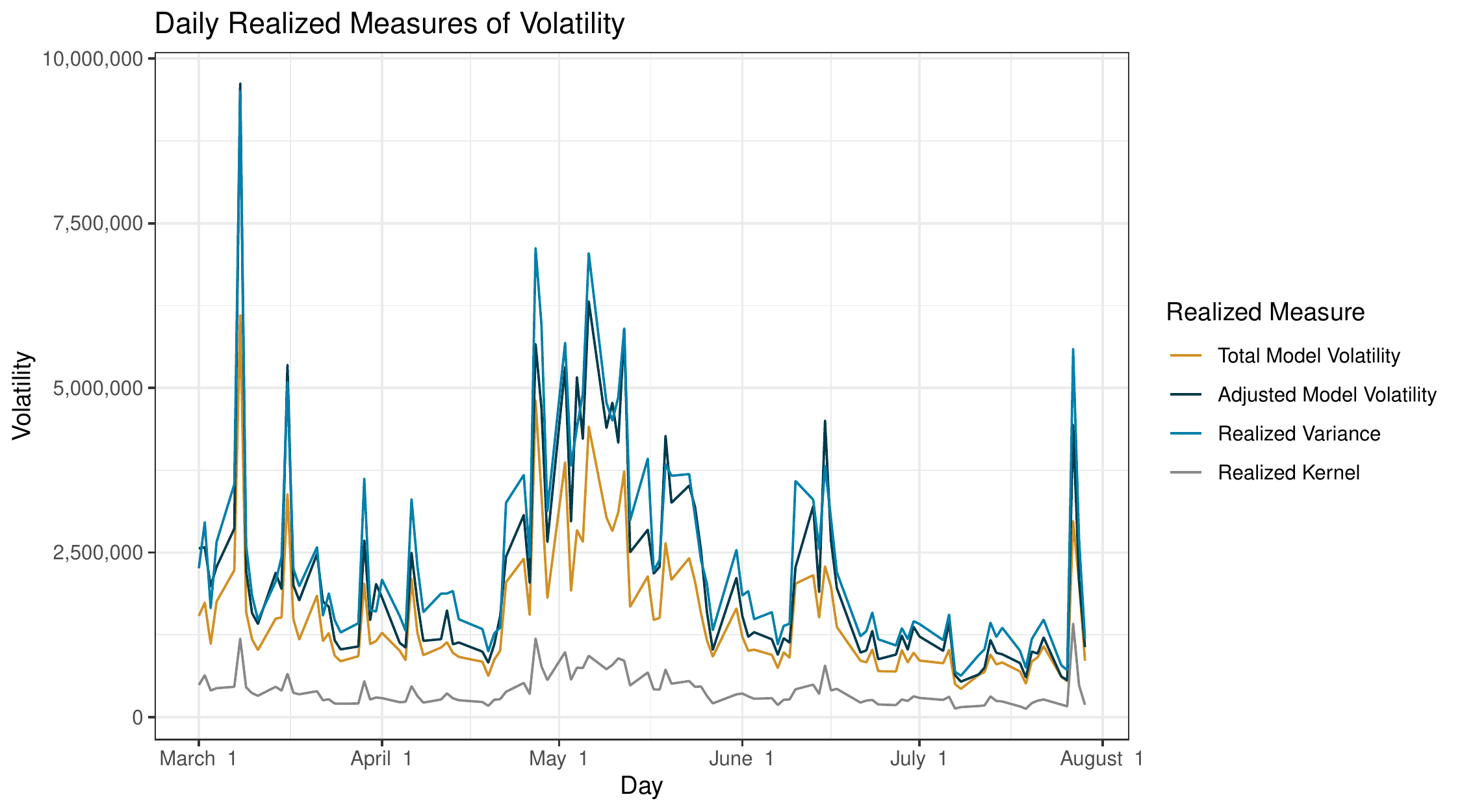}
\caption{The daily values of various volatility realized measures. The results are for the MA stock.}
\label{fig:realizedMA}
\end{figure}

\begin{figure}
\centering
\includegraphics[width=15cm]{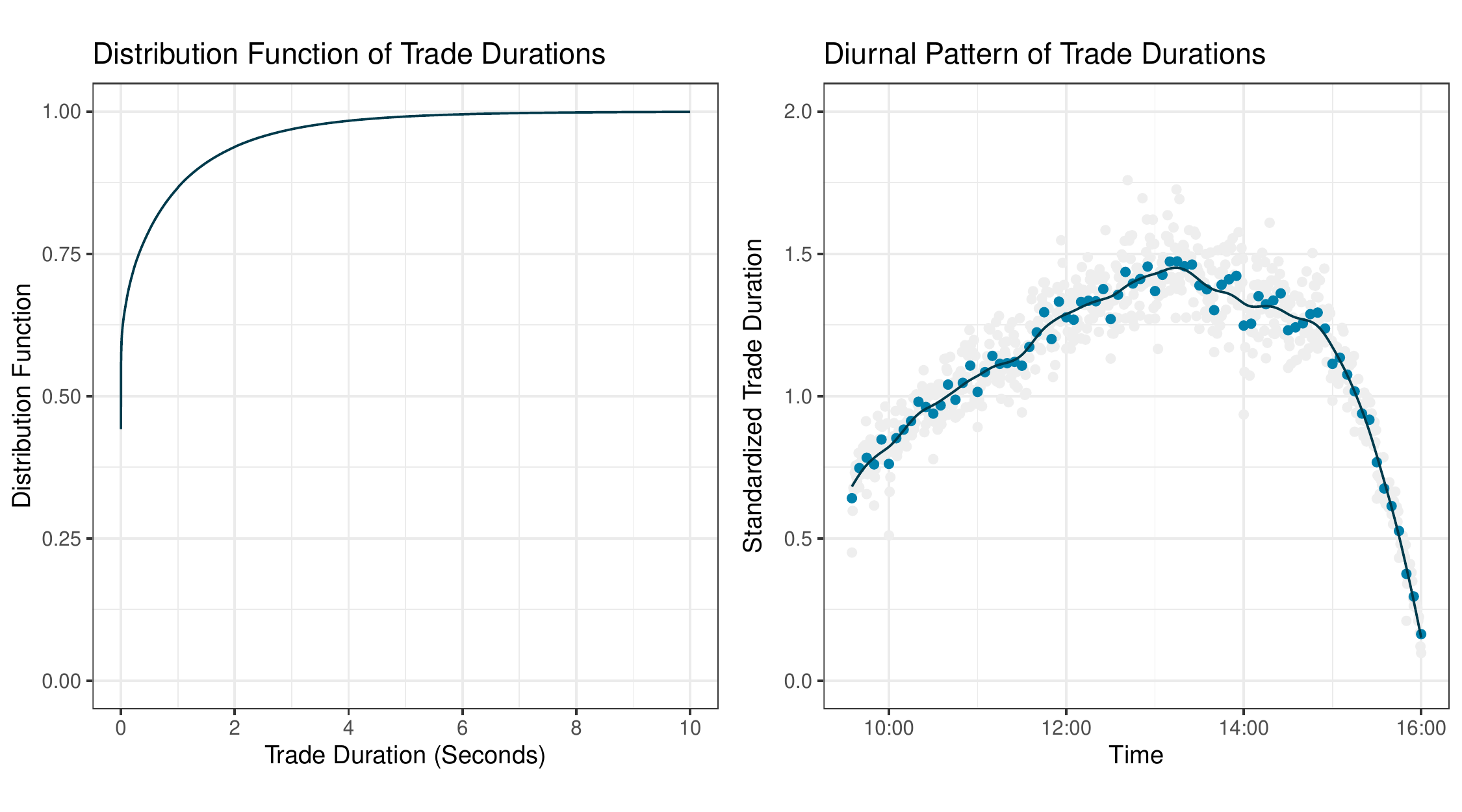}
\caption{Left: The empirical distribution function of trade durations. Right: The average trade durations in 5 minute (blue dots) and 30 second (grey dots) intraday intervals with a smoothed curve $\hat{f}_{\mathrm{dur}}(t_i)$. The results are for the MCD stock.}
\label{fig:durationMCD}
\end{figure}

\begin{figure}
\centering
\includegraphics[width=15cm]{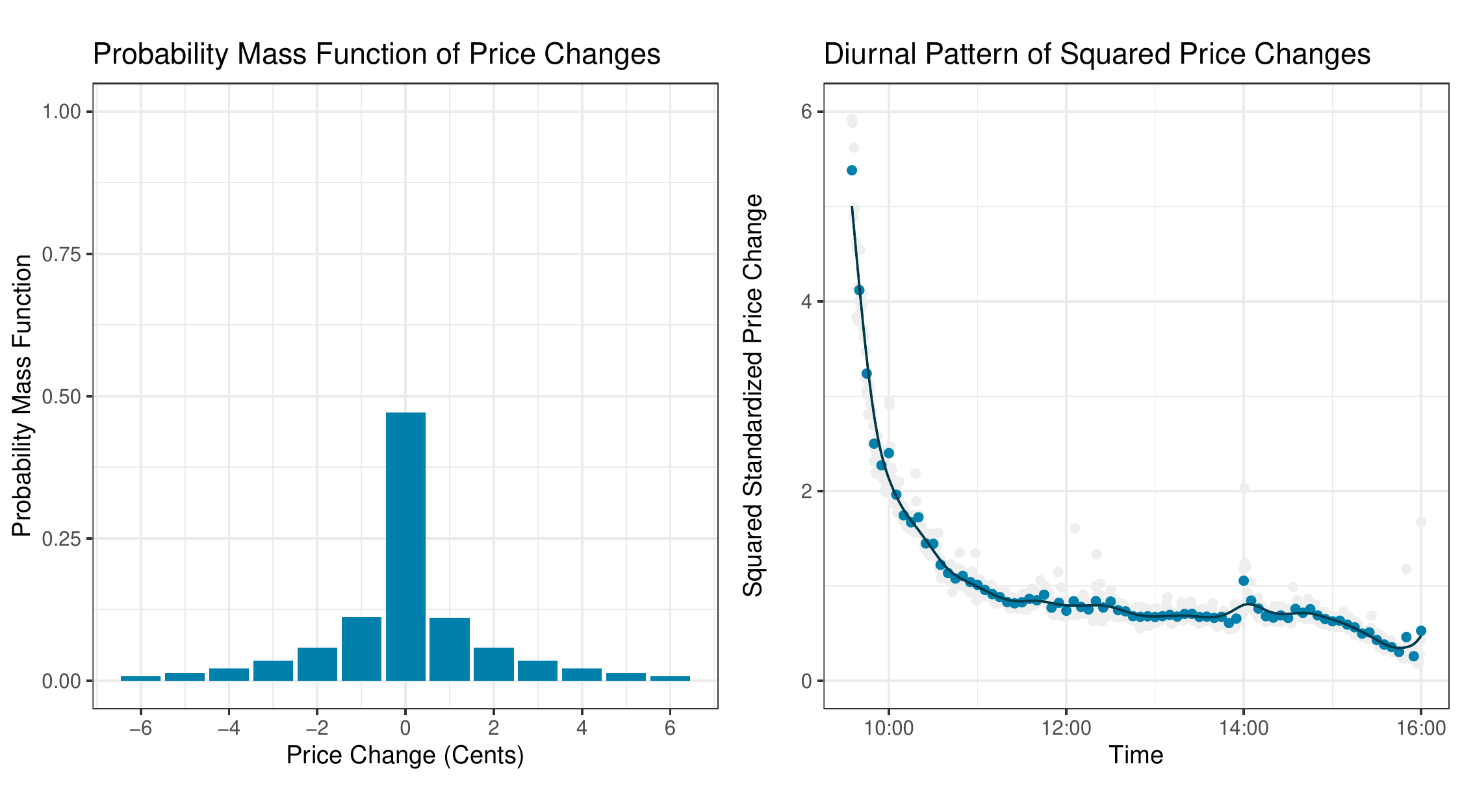}
\caption{Left: The empirical probability mass function of price changes. Right: The average squared price changes in 5 minute (blue dots) and 30 second (grey dots) intraday intervals with a smoothed curve $\hat{f}_{\mathrm{var}}(t_i)$. The results are for the MCD stock.}
\label{fig:returnMCD}
\end{figure}

\begin{figure}
\centering
\includegraphics[width=15cm]{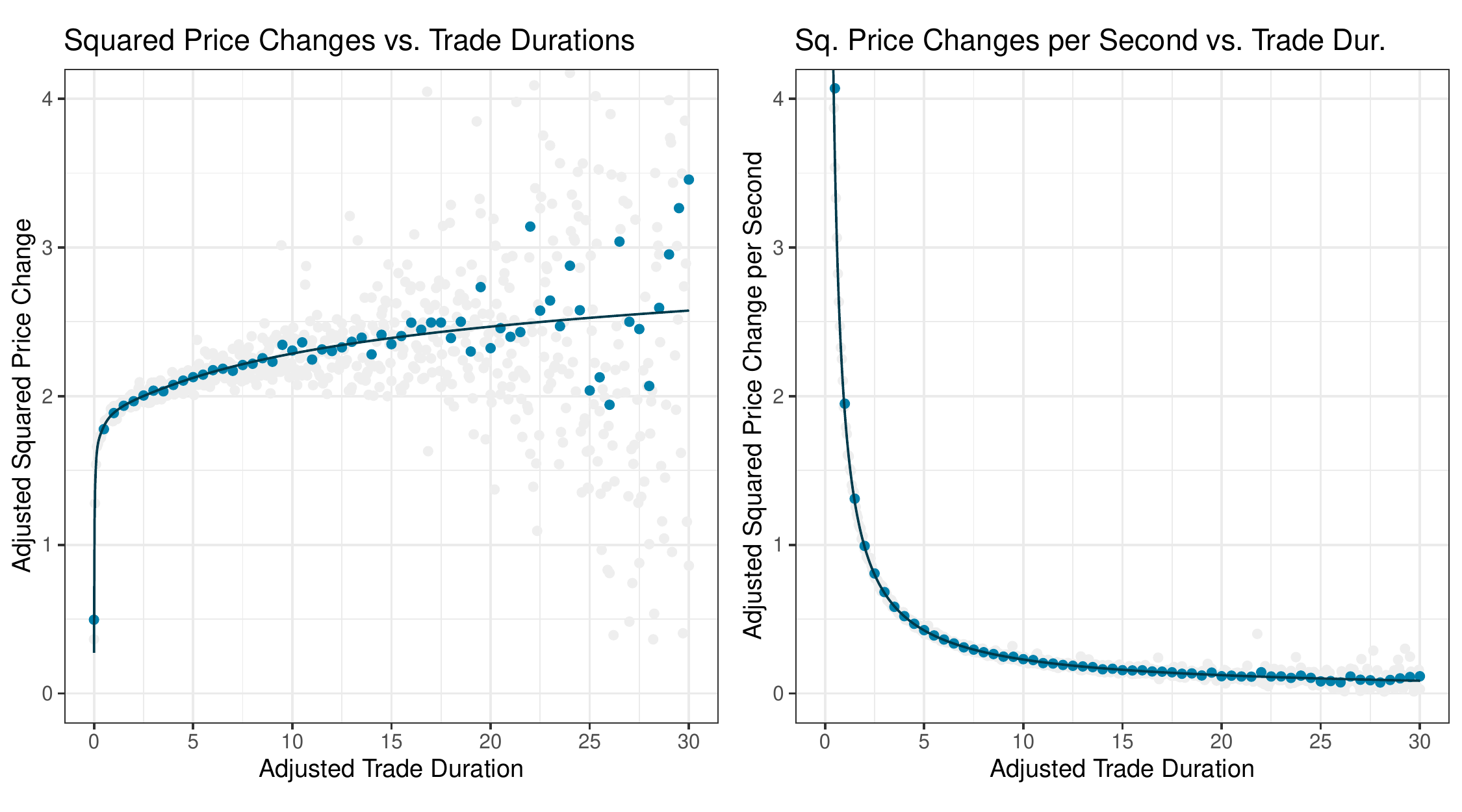}
\caption{Left: The average diurnally adjusted squared price changes in half second (blue dots) and 50 millisecond (grey dots) intervals of diurnally adjusted trade durations with a smoothed curve $\hat{f}_{\mathrm{rel}}(\tilde{d}_i)$. Right: The average diurnally adjusted squared price changes per second in half second (blue dots) and 50 millisecond (grey dots) intervals of diurnally adjusted trade durations with a smoothed curve $\hat{f}_{\mathrm{rel}}(\tilde{d}_i) / \tilde{d}_i$. The results are for the MCD stock.}
\label{fig:relationMCD}
\end{figure}

\begin{table}
\centering
\caption{The minimum, median, and maximum values of the estimated parameters of the models estimated on daily basis. The results are for the MCD stock.}
\label{tab:coefMCD}
\footnotesize
\begin{tabular}{llrrrrr}
\toprule
& & \multicolumn{5}{c}{Model} \\
\cmidrule(l{3pt}r{3pt}){3-7}
Coef. & Trans. & \multicolumn{1}{c}{Naive} & \multicolumn{1}{c}{No Infl.} & \multicolumn{1}{c}{Static Disp.} & \multicolumn{1}{c}{Static Mean} & \multicolumn{1}{c}{Proposed} \\
\midrule
          & Min &  & -0.385 & -0.587 &  & -0.504 \\ 
$\theta$  & Med &  & -0.306 & -0.457 &  & -0.408 \\ 
          & Max &  & -0.237 & -0.370 &  & -0.317 \\ \\
          & Min & 1.017 & 0.663 & 0.882 & 1.127 & 0.890 \\ 
$\omega$  & Med & 1.466 & 1.102 & 1.399 & 1.595 & 1.340 \\ 
          & Max & 2.556 & 2.247 & 2.608 & 2.744 & 2.546 \\ \\
          & Min &  & 0.894 &  & 0.821 & 0.938 \\ 
$\varphi$ & Med &  & 0.952 &  & 0.946 & 0.972 \\ 
          & Max &  & 0.998 &  & 0.997 & 0.999 \\ \\
          & Min &  & 0.026 &  & 0.029 & 0.019 \\ 
$\alpha$  & Med &  & 0.216 &  & 0.217 & 0.213 \\ 
          & Max &  & 0.279 &  & 0.461 & 0.301 \\ \\
          & Min &  &  & 0.184 & 0.144 & 0.160 \\ 
$\pi$     & Med &  &  & 0.241 & 0.203 & 0.214 \\ 
          & Max &  &  & 0.333 & 0.310 & 0.307 \\           
\bottomrule
\end{tabular}
\end{table}

\begin{table}
\centering
\caption{The R$^2$ statistics of the residuals and the squared residuals regressed on their lagged values with the average log-likelihood of an observation for the models estimated on daily basis. The results are for the MCD stock.}
\label{tab:fitMCD}
\footnotesize
\begin{tabular}{llrrrrr}
\toprule
& & \multicolumn{5}{c}{Model} \\
\cmidrule(l{3pt}r{3pt}){3-7}
Statistic & Lag & \multicolumn{1}{c}{Naive} & \multicolumn{1}{c}{No Infl.} & \multicolumn{1}{c}{Static Disp.} & \multicolumn{1}{c}{Static Mean} & \multicolumn{1}{c}{Proposed} \\
\midrule
               &   1 & 0.123 & 0.005 & 0.003 & 0.095 & 0.004 \\ 
AR R$^2$       &  10 & 0.162 & 0.011 & 0.005 & 0.122 & 0.006 \\ 
               & 100 & 0.166 & 0.013 & 0.008 & 0.124 & 0.008 \\ \\
               &   1 & 0.098 & 0.001 & 0.003 & 0.014 & 0.000 \\ 
ARCH R$^2$     &  10 & 0.147 & 0.002 & 0.041 & 0.015 & 0.002 \\ 
               & 100 & 0.190 & 0.007 & 0.074 & 0.021 & 0.005 \\    \\            
Log-Likelihood &     & -1.941 & -1.812 & -1.795 & -1.814 & -1.762 \\ 
\bottomrule
\end{tabular}
\end{table}

\begin{table}
\centering
\caption{The R$^2$ statistics of the residuals and the squared residuals regressed on their lagged values with the average log-likelihood of an observation for the alternative models estimated on daily basis. The results are for the MCD stock.}
\label{tab:altMCD}
\footnotesize
\begin{tabular}{llrrrrr}
\toprule
& & \multicolumn{5}{c}{Model} \\
\cmidrule(l{3pt}r{3pt}){3-7}
Statistic & Lag & \multicolumn{1}{c}{Var. Param.} & \multicolumn{1}{c}{GAS Mean} & \multicolumn{1}{c}{GAS Infl.} & \multicolumn{1}{c}{Normal} & \multicolumn{1}{c}{Student's-t} \\
\midrule
               &   1 & 0.050 & 0.037 & 0.003 & 0.008 & NA \\ 
AR R$^2$       &  10 & 0.067 & 0.045 & 0.005 & 0.016 & NA \\ 
               & 100 & 0.069 & 0.047 & 0.007 & 0.018 & NA \\ \\
               &   1 & 0.008 & 0.004 & 0.000 & 0.002 & NA \\ 
ARCH R$^2$     &  10 & 0.010 & 0.008 & 0.002 & 0.002 & NA \\ 
               & 100 & 0.018 & 0.013 & 0.005 & 0.007 & NA \\ \\
Log-Lik.       &     & -1.788 & -1.793 & -1.754 & -1.880 & -4.209 \\ 
\bottomrule
\end{tabular}
\end{table}

\begin{table}
\centering
\caption{The average log-likelihood, mean absolute error (MAE), and root mean squared error (RMSE) in the test sample for the naive and proposed models estimated on daily basis. The results are for the MCD stock.}
\label{tab:fcstMCD}
\footnotesize
\begin{tabular}{llrrrr}
\toprule
& & \multicolumn{2}{c}{Full Train Sample} & \multicolumn{2}{c}{Agg. Train Sample} \\
\cmidrule(l{3pt}r{3pt}){3-4} \cmidrule(l{3pt}r{3pt}){5-6}
Test Sample & Statistic & \multicolumn{1}{c}{Naive} & \multicolumn{1}{c}{Proposed} & \multicolumn{1}{c}{Naive} & \multicolumn{1}{c}{Proposed} \\
\midrule
           & Log-Lik. & -1.963 & -1.766 & -2.037 & -1.786 \\ 
Full       & MAE      &  1.405 &  1.432 &  1.405 &  1.440 \\ 
           & RMSE     &  2.566 &  2.346 &  2.566 &  2.333 \\ \\
           & Log-Lik. & -2.522 & -2.252 & -2.399 & -2.228 \\ 
Aggregated & MAE      &  2.103 &  2.064 &  2.103 &  2.067 \\ 
           & RMSE     &  3.246 &  2.963 &  3.246 &  2.946 \\              
\bottomrule
\end{tabular}
\end{table}

\begin{figure}
\centering
\includegraphics[width=15cm]{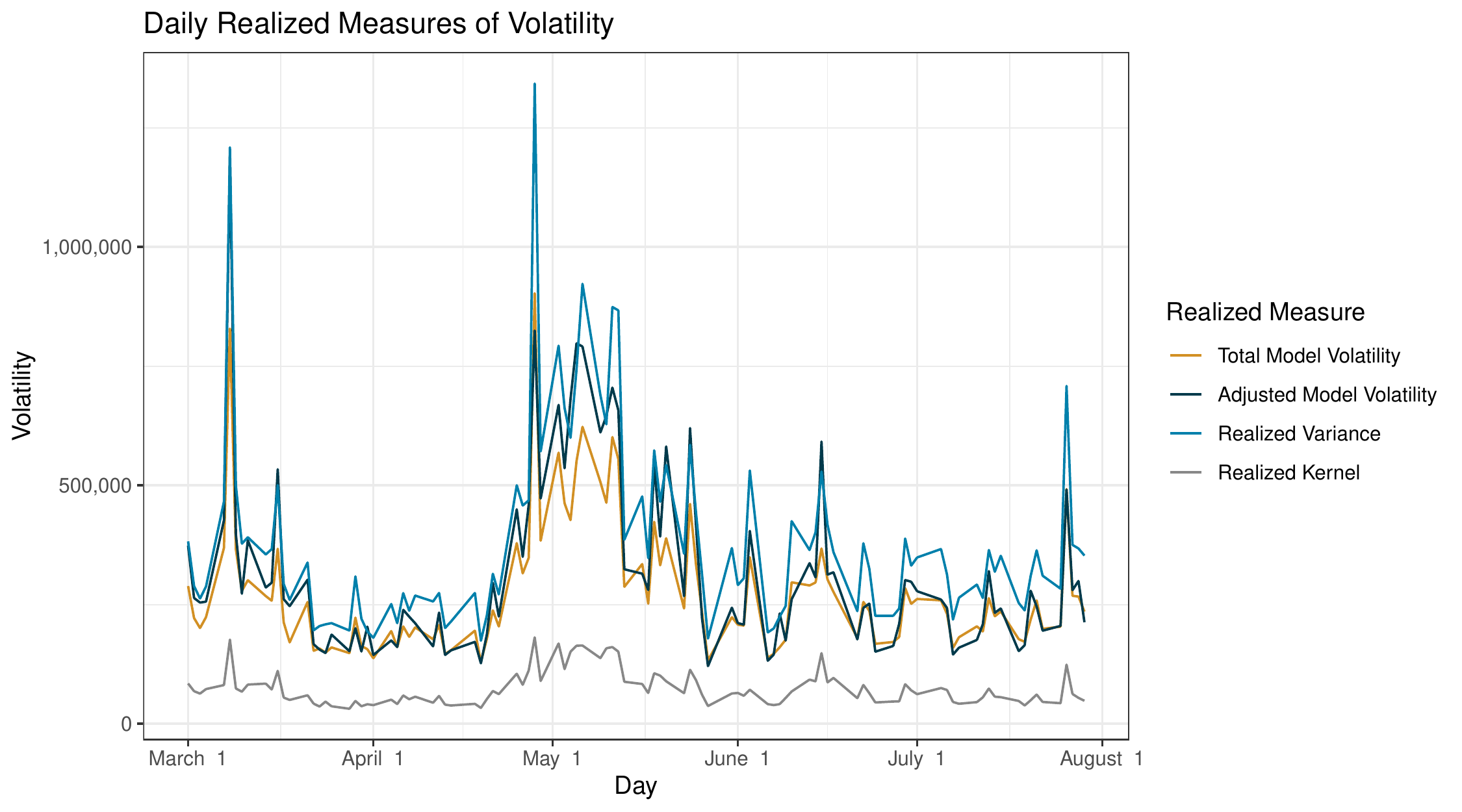}
\caption{The daily values of various volatility realized measures. The results are for the MCD stock.}
\label{fig:realizedMCD}
\end{figure}

\end{document}